\documentclass[twocolumn,showpacs,showkeys]{revtex4}
\usepackage{graphicx}
\usepackage{bm}
\usepackage{color}
\usepackage{amsmath}
\usepackage{natbib}

\begin{document}

\title{Hydrodynamic and kinetic models for spin-1/2 electron-positron quantum plasmas: Annihilation interaction, helicity conservation, and wave dispersion in magnetized plasmas}

\author{Pavel A. Andreev}
\email{andreevpa@physics.msu.ru}
\affiliation{Faculty of physics, Lomonosov Moscow State University, Moscow, Russian Federation.}

\date{\today}

\begin{abstract}
We discuss complete theory of spin-1/2 electron-positron quantum plasmas, when electrons and positrons move with velocities mach smaller than the speed of light. We derive a set of two fluid quantum hydrodynamic equations consisting of the continuity, Euler, spin (magnetic moment) evolution equations for each species. We explicitly include the Coulomb, spin-spin, Darwin and annihilation interactions. The annihilation interaction is the main topic of the paper. We consider contribution of the annihilation interaction in the quantum hydrodynamic equations and in spectrum of waves in magnetized electron-positron plasmas. We consider propagation of waves parallel and perpendicular to an external magnetic field. We also consider oblique propagation of longitudinal waves. We derive set of quantum kinetic equations for electron-positron plasmas with the Darwin and annihilation interactions. We apply the kinetic theory for the linear wave behavior in absence of external fields. We calculate contribution of the Darwin and annihilation interactions in the Landau damping of the Langmuir waves. We should mention that the annihilation interaction does not change number of particles in the system. It does not related to annihilation itself, but it exists as a result of interaction of an electron-positron pair via conversion of the pair into virtual photon. A pair of the non-linear Schrodinger equations for electron-positron plasmas including the Darwin and annihilation interactions. Existence of conserving helicity in electron-positron quantum plasmas of spinning particles with the Darwin and annihilation interactions is demonstrated. We show that annihilation interaction plays an important role in quantum electron-positron plasmas giving contribution of the same magnitude as the spin-spin interaction.
\end{abstract}

\pacs{52.27.Ep, 52.30.Ex, 52.25.Xz, 52.35.We}% PACS, the Physics and Astronomy
                             % Classification Scheme.
%Use showkeys class option if keyword

\maketitle

%%%%%%%%%%TEXT

%52.27.Ep   Electron-positron plasmas

%52.27.Lw Dusty or complex plasmas; plasma crystals

%52.30.Ex   Two-fluid and multi-fluid plasmas
%52.35.Fp   Electrostatic waves and oscillations (e.g., ion-acoustic waves)

%52.35.Hr   Electromagnetic waves (e.g., electron-cyclotron, Whistler, Bernstein, upper hybrid, lower hybrid)

%52.35.Dm   Sound waves

%52.25.Xz Magnetized plasmas
%52.25.Ya Neutrals in plasmas

%52.35.Bj Magnetohydrodynamic waves (e.g., Alfven waves)
%52.25.Dg Plasma kinetic equations

%52.35.We   Plasma vorticity
%52.55.Wq   Current drive; helicity injection
%67.10.Db   Fermion degeneracy

\section{Introduction}

In classical plasmas the model of electron-positron plasmas does not differ from model of electron-ion plasmas. Electrons, positrons, and ions are involved in the electromagnetic interaction between charges. Some interesting effects exist in classic electron-plasmas due to the equality of masses and module of charges of both species. Considering quantum plasmas we include spin of particles \cite{Maksimov Izv 2000}-\cite{Andreev Asenjo 13}. Quantum nature of particles requires us to include the Darwin interaction as well \cite{Ivanov Darwin}, \cite{Asenjo NJP 12}. This interaction has weakly-relativistic (semi-relativistic) nature \cite{Landau 4}, hence it is neglected in most of the papers on quantum plasmas. Quantum model of electron-positron requires to consider extra interaction, which does not exist between electrons or between ions. This is, so called, the annihilation interaction \cite{Pirenne 1947}-\cite{Berestetski ZETP 49 b}, see also Ref. \cite{Landau 4} section 83. It is not annihilation of electron-positron pairs itself. However it is the semi-relativistic trace of electron-positron interaction, when an electron-positron pair transforms in the virtual photon, which splits back into the electron-positron pair.

Studying of the annihilation interaction in electron-positron quantum plasmas is an essential part of understanding of the quantum and relativistic properties of plasmas \cite{Ruyer PP 13}-\cite{Uzdensky arxiv review 14}.

The quantum hydrodynamic model for electron-plasmas with the annihilation interaction is one of main topics of the paper. We derive the model and we use it to study spectrum of small amplitude collective excitations in magnetized electron-positron plasmas.

We should mention that classical relativistic properties of electron-positron plasmas, as well as quantum properties of electron-positron and electron-positron-ion plasmas have been under consideration in last years. For instance thermal-inertial effects on magnetic reconnection in relativistic pair plasmas \cite{Luca Comisso arXiv 14},
self-modulation of nonlinear waves in a weakly magnetized relativistic electron-positron plasma with temperature \cite{Asenjo PRE 12},                   and nonlinear Alfven waves in a strongly magnetized relativistic electron-positron plasma \cite{Lopez PRE 13}
have been studied on the path of research of classic relativistic electron-positron plasmas. In these papers a hydrodynamic model of relativistic quantum plasmas with temperature \cite{Mahajan 03} was applied.

Before speaking of quantum properties of electron-positron plasmas, we should present a brief description of the field of quantum plasmas of spinning particles. Method of description of quantum plasmas was developed in 1999-2001 in Refs. \cite{Maksimov Izv 2000}, \cite{MaksimovTMP 2001}, \cite{MaksimovTMP 1999}. This method arises as a representation of the many-particle Schrodinger equation with the charge-charge Coulomb and spin-spin interactions in terms of collective variables. The collective variables are the microscopic observable variables suitable for description of many-particle systems. They are the particle concentration $n$, the momentum density $\textbf{j}=n\textbf{v}$, the pressure $p$, the magnetic moment (spin) density $\mbox{\boldmath $\mu$}$, the energy density $\varepsilon$, the spin-current (magnetization flux) $J^{\alpha\beta}$, etc. These variables are determined via the many-particle wave function, or wave spinor for spinning particles. Evolution of the wave function obeys the many-particle Schrodinger equation. Hence, applying the Schrodinger equation we can derive equations of evolution of collective variables. The particle number evolution (the continuity equation), the momentum balance equation (the Euler equation), the energy balance equation, the magnetic moment evolution equation were derived for many-particle systems in 2000-2001. This set equation is a generalization of the five moment approximation ($n$, $\textbf{v}$, and $\varepsilon$) for spinning particles appearing as the eight-moment approximation ($n$, $\textbf{v}$, $\mbox{\boldmath $\mu$}$, and $\varepsilon$). This set of equations in not closed set of equations. It is a long chain of equations, which should be truncated. Truncation is making of an approximation, but this is also an "explanation" to our method of properties of the system under consideration. Generalization of the eight-moment approximation were developed in Ref. \cite{Andreev spin current}, where the spin-current evolution equation were derived to get richer information on the spin evolution properties. Terms describing interparticle interaction in hydrodynamic equations contains two-particle functions containing two-particle correlations. Main attention of Refs. \cite{Maksimov Izv 2000}, \cite{MaksimovTMP 2001}, \cite{Andreev RPJ 07}, \cite{MaksimovTMP 1999}, \cite{Andreev spin current}, \cite{Andreev PRB 11} were focused on the self-consistent field approximation, when two-particle functions appears as products of two corresponding one-particle functions. Nevertheless, quantum correlations related to the exchange Coulomb and spin-spin interactions were also considered in Refs. \cite{MaksimovTMP 2001}, \cite{MaksimovTMP 1999}.

Since then various wave phenomenon have been studied for spin-1/2 quantum plasmas \cite{Andreev VestnMSU 2007}, \cite{Marklund PRL07}, \cite{Andreev AtPhys 08}, \cite{Andreev Asenjo 13}, \cite{Maksimov VestnMSU 2000}-\cite{Asenjo PL A 12}. Presence of spin induces the spin-spin interaction via the magnetic field created by magnetic moments \cite{MaksimovTMP 2001}. Spin also causes the spin-current interaction, i.e. interaction of the magnetic moments and electric currents via the magnetic field. These interactions change dispersion of plasma waves in compare with the spinless case. Spin evolution leads to extra waves in plasmas. These new spin-plasma waves were found in several papers, see Refs. Vagin et al \cite{Vagin Izv RAN 06}, Andreev and Kuz'menkov \cite{Andreev VestnMSU 2007}, Brodin et al \cite{Brodin PRL 08} for the spin-plasma waves propagating perpendicular to an external magnetic field, see Refs. Misra et al \cite{Misra JPP 10}, Andreev and Kuz'menkov \cite{Andreev IJMP 12}, \cite{pavelproc} for the spin-plasma waves propagating parallel to an external magnetic field, spin waves in quantum plasmas propagating due to perturbations of magnetic field with no contribution of electric field in the wave propagation were considered in Refs. \cite{Andreev VestnMSU 2007}, \cite{Andreev IJMP 12}, new branches of wave related to spin-current evolution were found in Refs. \cite{Andreev spin current} and \cite{Trukhanova 1403}. Equality of masses of electrons and positrons changes dispersion dependencies of spin-plasma waves. It happens with usual, spinless, plasma wave. It also affects plasmas of spinning particles.

Let us also mention that the Karpman-Washimi magnetization and the Karpman-Washimi interaction for plasmas of spinning particles were considered in Ref. \cite{Dae PP 13}.

An interesting application of quantum hydrodynamics to vorticity of spinning particles was suggested in Refs. \cite{Mahajan PRL 11} and \cite{Braun PRL 12}.

Reviews of some topics studied for quantum plasmas are presented in Refs. \cite{Shukla UFN 10} and \cite{Shukla RMP 11}.

Electron-positron and electron-positron-ion spin-1/2 quantum plasmas and their wave properties have been under consideration in recent years \cite{Bains PP 10}-\cite{Brodin PRL 10}.

Hydrodynamic representation of the Schrodinger equation for a single particle in an external field was made by Madelung in 1926 \cite{Madelung}. Equations obtained by Madelung for a single particle look similar to hydrodynamic equations for many-particle systems of classic particles. This similarity was used by Rand in 1964 \cite{Rand PF 64} for quantum hydrodynamic description of quantum plasmas. The method of many-particle quantum hydrodynamics suggested by Kuz'menkov et al 1999-2001 \cite{MaksimovTMP 2001}, \cite{MaksimovTMP 1999} differs from the single particle one by the fair treating of the many-particle quantum dynamics. That opens a lot of possibilities for consideration of different physical systems (see for instance \cite{Andreev PRB 11} and \cite{Andreev PRA08}).

This paper is organized as follows. Quantum hydrodynamic model for electron-positron quantum plasmas of spinning particles is developed in Sec. II. In Sec. III we consider dispersion of waves propagating parallel to the external magnetic
field. We describe contribution of the annihilation interaction along with the spin-spin interaction in longitudinal and transverse waves including spin-plasma waves. We present contribution of the Darwin and exchange interactions in the longitudinal Langmuir wave. In Sec. IV we consider dispersion of waves propagating perpendicular to the external magnetic
field. In Sec. V we pay special attention to the longitudinal waves. We consider oblique propagation of the longitudinal waves. In Sec. VI we present generalization of the theory developed in section II. From the first principles we develop set of kinetic equations for spinning electrons and positrons with the annihilation interaction. We apply kinetic theory to calculation of the Landau damping for the quantum Langmuir waves. In Sec. VII we present a pair of the non-linear Schrodinger equations for electron-positron plasmas including the Darwin and annihilation interactions. In Sec. VIII we derive equations for the vorticity evolution and show existence of conserving helicity in electron-positron quantum plasmas of spinning particles with the Darwin and annihilation interactions. In Sec. IX brief summary of obtained results is presented.

\section{Background of the Model}

In this paper we use and develop the method of many-particle quantum hydrodynamics. The method was suggested by Kuz'menkov and Maximov in 1999 \cite{MaksimovTMP 1999}. It was done for spinless charged particles. Generalization of this method for spin-1/2 charged particles was presented by Kuz'menkov et al. in 2000-2001 \cite{Maksimov Izv 2000}, \cite{MaksimovTMP 2001}. Main idea of the method is the representation of the many-particle Schrodinger equation in terms of collective observable variables suitable for description of quantum plasmas or other physical systems. Hence the many-particle Schrodinger equation
\begin{equation}\label{E-P_QP Schrodinger eq} \imath\hbar\partial_{t}\psi(R,t)=\hat{H}\psi(R,t) \end{equation}
is the starting point of our paper. In equation (\ref{E-P_QP Schrodinger eq}) we use the following quantities: $\hbar$ is the reduced Plank constant, $\partial_{t}$ is the time derivative, $\psi(R,t)$ is the many-particle wave function, $R=(\textbf{r}_{1}, \textbf{r}_{2}, ..., \textbf{r}_{N})$ is the set of coordinates of $N$ particles, with $N$ is the full number of particles in the system, $\hat{H}$ is the Hamiltonian for a system under consideration. We deal with the spinning particles, consequently, $\psi(R,t)$ is the spinor function.

To describe system of electrons and positrons we present the Hamiltonian for the many-particle Schrodinger equation (\ref{E-P_QP Schrodinger eq}). Explicit form of the Hamiltonian allows to derive a set of quantum hydrodynamic (QHD) equations. Method of derivation is described in Refs. \cite{MaksimovTMP 2001}, \cite{Andreev Asenjo 13}, \cite{MaksimovTMP 1999}, \cite{Andreev PRB 11}, \cite{Andreev PRA08}. Presenting the QHD equations below, we give some tips for their derivation as well. Microscopic Hamiltonian for electron-positron plasmas of particles moving with velocities mach smaller than the speed of light appears as
$$\hat{H}=\sum_{i=1}^{N}\biggl(\frac{1}{2m_{i}}\hat{\textbf{D}}_{i}^{2}+q_{i}\varphi^{ext}_{i}-\gamma_{i}\mbox{\boldmath $\sigma$}_{i}\textbf{B}_{i(ext)}\biggr)$$
$$+\frac{1}{2}\sum_{i,j\neq i}^{N}(e_{i}e_{j}G_{ij}-\gamma_{i}\gamma_{j}G^{\alpha\beta}_{ij}\sigma^{\alpha}_{i}\sigma^{\beta}_{j})+$$
$$-\frac{1}{2}\sum_{i,j\neq i}^{N}\frac{\pi e_{i}e_{j}\hbar^{2}}{m^{2}c^{2}}\delta(\textbf{r}_{i}-\textbf{r}_{j})$$
\begin{equation}\label{E-P_QP Hamiltonian}-\sum_{i=1}^{N_{e-}}\sum_{j=N_{e-}+1}^{N_{e-}+N_{e+}}\frac{\pi e_{i}e_{j}\hbar^{2}}{2m^{2}c^{2}}(3+\mbox{\boldmath $\sigma$}_{i}\mbox{\boldmath $\sigma$}_{j})\delta(\textbf{r}_{i}-\textbf{r}_{j})\end{equation}
where $N$ is the full number of particles, $N_{e-}$, $N_{e+}$ are numbers of electrons and positrons correspondingly, and $N=N_{e-}+N_{e+}$, $i$ is a number of particle, $m_{i}$ is the mass of particle with number $i$, below we consider system of particles with equal masses, $q_{i}$ is the charge of particle, $\gamma_{i}$ is the gyromagnetic ratio, for electrons and positrons it can be written as $\gamma_{i}\approx1.00116\mu_{B}$, $\mu_{B}=q_{i}\hbar/(2m_{i}c)$ is the Bohr magneton, the difference of $\mid\gamma_{e}\mid$ from the Bohr magneton includes contribution of the anomalous magnetic dipole moment, $\varphi^{ext}_{i}$ is the scalar potential of an external electromagnetic field acting on particle with number $i$, $\textbf{B}_{i(ext)}$ is the external magnetic field,
$(\hat{\textbf{D}}_{i}\psi)(R,t)=((-\imath\hbar\nabla_{i}-\frac{q_{i}}{c}\textbf{A}_{i,ext})\psi)(R,t)$, with $\textbf{A}_{i,ext}$ is the vector potential of an external electromagnetic field acting on particle, $\mbox{\boldmath $\sigma$}_{i}$ is the Pauli matrixes describing spin of particles, they satisfy the following commutation relation
$[\sigma^{\alpha}_{i},\sigma^{\beta}_{j}]=2\imath\delta_{ij}\varepsilon^{\alpha\beta\gamma}\sigma^{\gamma}_{i} $, $\delta_{ij}$ is the Kroneckers delta, $\varepsilon^{\alpha\beta\gamma}$ is the Levi-Civita symbol,
$\textbf{B}_{i(ext)}=\nabla_{i}\times \textbf{A}_{i(ext)}$,
$\textbf{E}_{i(ext)}=-\nabla_{i}\varphi_{ext}(\textbf{r}_{i},t)-\frac{1}{c}\partial_{t}\textbf{A}_{ext}(\textbf{r}_{i},t)$, $G_{pn}=\frac{1}{r_{ij}}$ is the Green function of the Coulomb interaction containing module of the interparticle distance $\textbf{r}_{ij}=\textbf{r}_{i}-\textbf{r}_{j}$, and
\begin{equation}\label{E-P_QP} G^{\alpha\beta}_{ij}=4\pi\delta_{\alpha\beta}\delta(\textbf{r}_{ij})+\nabla^{\alpha}_{i}\nabla^{\beta}_{i}\frac{1}{r_{ij}}\end{equation}
is the Green function of the spin-spin interaction.

\begin{figure}
\includegraphics[width=8cm,angle=0]{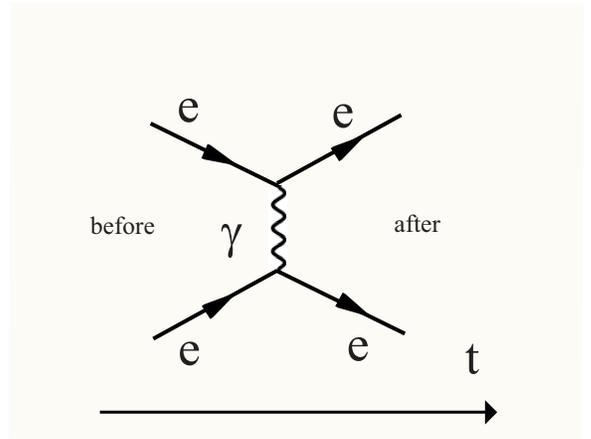}
\caption{\label{FD_ee} The figure shows the Feynman diagrams for electron-electron interaction. Long arrow shows direction of time evolution.}
\end{figure}

\begin{figure}
\includegraphics[width=8cm,angle=0]{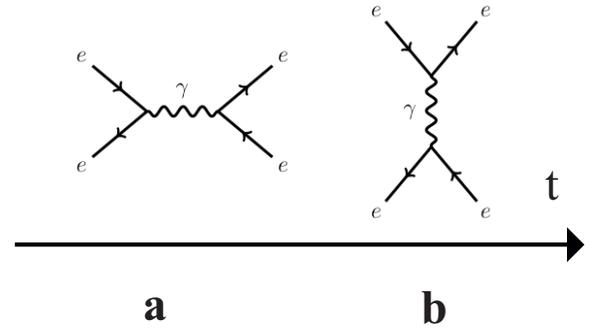}
\caption{\label{FD_ep} The figure shows the Feynman diagrams for electron-positron interaction. Long arrow shows direction of time evolution. Small arrows pointing in direction of time present electrons. Small arrows pointing in opposite direction describes positrons.}
\end{figure}

Let us describe physical meaning of different terms in the Hamiltonian (\ref{E-P_QP Hamiltonian}). The first term describes the kinetic energy. The second term is the potential energy of charges in the external electromagnetic field. The third term is the potential energy of magnetic moments in the external magnetic field. These three terms are related to motion of each particle in the external electromagnetic field. Other terms describe interparticle interactions: the Coulomb, spin-spin, Darwin (see \cite{Landau 4} sections 33 and 83 \textit{and} \cite{Strange book} formula (4.74b)), annihilation interactions, correspondingly (for annihilation interaction see Refs. \cite{Pirenne 1947}, \cite{Berestetski ZETP 49 a}, \cite{Berestetski ZETP 49 b}, and textbook \cite{Landau 4} section 83).

The annihilation interaction plays an essential role at description of spectrum of positronium (bound atom-like state of an electron and a positron) \cite{Lewis PRA 73}.

Electron-electron interaction is described by diagram presented on
Fig.(\ref{FD_ee}). Fig. (\ref{FD_ep}.b) shows a Feynman diagram
describing the Coulomb, Darwin, spin-spin and other interactions
including in the Breit Hamiltonian \cite{Landau 4} (see section
83). Usual spin-spin interaction appears from diagram on Fig.
(\ref{FD_ep}.b). The annihilation interaction gives extra term
depending on spins of interacting particles. Hence it modifies
spin-spin interaction. Full annihilation interaction including
shift of the spin-spin interaction appears from diagram on Fig.
(\ref{FD_ep}.a). Both effects have same magnitude
$\sim\mu_{B}^{2}$. We point it out to notice that the annihilation
interaction needs to be included in hydrodynamic equations.

We derive two fluid quantum hydrodynamics for electron-positron plasmas. To this end we introduce the concentration for electrons
\begin{equation}\label{E-P_QP} n_{e}(\textbf{r},t)=\int
\psi^{+}(R,t)\sum_{i=1}^{N_{e-}}\delta(\textbf{r}-\textbf{r}_{i})\psi(R,t)dR,\end{equation}
and the concentration of positrons
\begin{equation}\label{E-P_QP} n_{p}(\textbf{r},t)=\int
\psi^{+}(R,t)\sum_{i=N_{e-}+1}^{N_{e-}+N_{e+}}\delta(\textbf{r}-\textbf{r}_{i})\psi(R,t)dR.\end{equation}
These two function are defined via the same many-particle wave function $\psi(R,t)$, which describes behavior of all particles in the system and obeys the Schrodinger equation (\ref{E-P_QP Schrodinger eq}).

\subsection{Continuity and Euler equations}

Differentiating concentration of particles of species $a$ and applying the Schrodinger equation with Hamiltonian (\ref{E-P_QP Schrodinger eq}) we find the
continuity equation
\begin{equation}\label{E-P_QP}\partial_{t}n_{a}(\textbf{r},t)+\nabla\textbf{j}_{a}(\textbf{r},t)=0, \end{equation}
where
$$\textbf{j}_{a}(\textbf{r},t)=\int dR\sum_{i}\delta(\textbf{r}-\textbf{r}_{i})\frac{1}{2m_{i}}\times$$
\begin{equation}\label{E-P_QP} \times\biggl((\textbf{D}_{i}\psi)^{+}(R,t)\psi(R,t)+h.c.\biggr)\end{equation}
is the particles current for species $a$.

Using the particle current we can introduce the velocity field for each species of particles
$\textbf{j}_{a}(\textbf{r},t)=n_{a}(\textbf{r},t)\textbf{v}_{a}(\textbf{r},t)$.

We have the continuity equations for electrons
\begin{equation}\label{E-P_QP continuity eq for electron} \partial_{t}n_{e}+\nabla(n_{e}\textbf{v}_{e})=0, \end{equation}
and positrons
\begin{equation}\label{E-P_QP continuity eq for positron} \partial_{t}n_{p}+\nabla(n_{p}\textbf{v}_{p})=0. \end{equation}

The annihilation interaction does not include real annihilation of electron-positron pairs. It involves virtual annihilation, if we speak in terms of the Feynman diagrams. Consequently the number of electrons and positrons do not change and the continuity equations have the traditional form (\ref{E-P_QP continuity eq for electron}) and (\ref{E-P_QP continuity eq for positron}).

We also have a couple of the Euler equations, one of them for electrons
$$mn_{e}(\partial_{t}+\textbf{v}_{e}\nabla)\textbf{v}_{e}+\nabla p_{e}-\frac{\hbar^{2}}{4m}n_{e}\nabla\Biggl(\frac{\triangle n_{e}}{n_{e}}-\frac{(\nabla n_{e})^{2}}{2n_{e}^{2}}\Biggr)$$
$$+\frac{\hbar^{2}}{4m\gamma^{2}_{e}}\partial^{\beta}\biggl(n_{e}(\partial^{\beta}\mu_{e}^{\gamma})\nabla\mu_{e}^{\gamma}\biggr)
=q_{e}n_{e}\textbf{E}$$
$$+\frac{q_{e}}{c}n_{e}[\textbf{v}_{e},\textbf{B}]+n_{e}\mu^{\beta}_{e}\nabla B^{\beta}$$
$$+\frac{\pi q_{e}^{2}\hbar^{2}}{m^{2}c^{2}}n_{e}\nabla n_{e}+\frac{\pi q_{e}q_{p}\hbar^{2}}{m^{2}c^{2}}n_{e}\nabla n_{p}$$
\begin{equation}\label{E-P_QP Euler equation for electrons}+\frac{3}{2}\frac{\pi q_{e}q_{p}\hbar^{2}}{m^{2}c^{2}}n_{e}\nabla n_{p}
+2\pi n_{e}\mu^{\beta}_{e}\nabla (n_{p}\mu^{\beta}_{p}),\end{equation}
and the second one for positrons
$$mn_{p}(\partial_{t}+\textbf{v}_{p}\nabla)\textbf{v}_{p}+\nabla p_{p}-\frac{\hbar^{2}}{4m}n_{p}\nabla\Biggl(\frac{\triangle n_{p}}{n_{p}}-\frac{(\nabla n_{p})^{2}}{2n_{p}^{2}}\Biggr)$$
$$+\frac{\hbar^{2}}{4m\gamma^{2}_{p}}\partial^{\beta}\biggl(n_{p}(\partial^{\beta}\mu_{p}^{\gamma})\nabla\mu_{p}^{\gamma}\biggr)
=q_{p}n_{p}\textbf{E}$$
$$+\frac{q_{p}}{c}n_{p}[\textbf{v}_{p},\textbf{B}]+n_{p}\mu^{\beta}_{p}\nabla B^{\beta}$$
$$+\frac{\pi q_{p}^{2}\hbar^{2}}{m^{2}c^{2}}n_{p}\nabla n_{p}+\frac{\pi q_{e}q_{p}\hbar^{2}}{m^{2}c^{2}}n_{p}\nabla n_{e}$$
\begin{equation}\label{E-P_QP Euler equation for positrons}+\frac{3}{2}\frac{\pi q_{e}q_{p}\hbar^{2}}{m^{2}c^{2}}n_{p}\nabla n_{e}
+2\pi n_{p}\mu^{\beta}_{p}\nabla (n_{e}\mu^{\beta}_{e}).\end{equation}
In these equations we have used the following notations: $p_{a}$ is the thermal pressure, or the Fermi pressure for degenerate systems of particles, $\mbox{\boldmath $\mu$}_{a}$ is proportional to the magnetization $\textbf{M}_{a}(\textbf{r},t)$ of species $a$: $\textbf{M}_{a}(\textbf{r},t)=n_{a}\mbox{\boldmath $\mu$}_{a}$, with the reduced magnetization.
The first group of terms in the left-hand side of equations (\ref{E-P_QP Euler equation for electrons}) and (\ref{E-P_QP Euler equation for positrons}) have kinematic nature and related to the local centre of mass motion, the second terms describe the contribution of pressure, related to particles motion in the frame of the local centre of mass, the third groups of terms are the spinless part of the quantum Bohm potential, the last terms in the left-hand sides of equations (\ref{E-P_QP Euler equation for electrons}) and (\ref{E-P_QP Euler equation for positrons}) are spin dependent parts of the quantum Bohm potential.

The quantum Bohm potential $\textbf{Q}$ is presented in the single-particle approximation. It means that these terms are simplified even in compare with the approximation of independent particles. For N independent fermions we find that different particles are in different quantum states describing with single particle wave functions. Hence the quantum Bohm potential appears as a sum over all occupied states $\textbf{Q}=\sum_{s}\biggl[-\frac{\hbar^{2}}{4m}\rho_{s}\nabla\biggl(\frac{\triangle\rho_{s}}{\rho_{s}}-\frac{(\nabla\rho_{s})^{2}}{2\rho_{s}^{2}}\biggr)$ $+\frac{\hbar^{2}}{4m}\partial^{\beta}\biggl(\rho_{s}(\partial^{\beta}\mu_{s}^{\gamma})\nabla\mu_{s}^{\gamma}\biggr)\biggr]$, where $\rho_{s}=\psi_{s}^{*}\psi_{s}$ and $\mbox{\boldmath $\mu$}_{s}=\psi_{s}^{*}\mbox{\boldmath $\sigma$}_{a}\psi_{s}$, $\psi_{s}$ is the single particle wave functions describing quantum state $s$.

The first term in the right-hand side of equations (\ref{E-P_QP Euler equation for electrons}) and (\ref{E-P_QP Euler equation for positrons}) are interaction of charges with the external and internal electric fields. The internal electric field exists due to the Coulomb interaction between particles, which is explicitly included in the basic Hamiltonian (\ref{E-P_QP Hamiltonian}). The second terms are the magnetic parts of the Lorentz force. The basic Hamiltonian (\ref{E-P_QP Hamiltonian}) contains interaction of moving charges with an external magnetic field only, but here we assume that full magnetic field, i.e. the field created by magnetic moments and currents in plasmas, is presented in the Euler equations. Action of the magnetic field created by the magnetic moments on the electric currents and action of the magnetic field created by the electric currents on the magnetic moments are the spin-current interaction. Action of the magnetic field created by the electric currents on the electric currents is the current-current interaction. The spin-current and the current-current interaction forming this term were considered in the following Refs., see \cite{Andreev RPJ 07} for the spin-current interaction, see \cite{Ivanov Darwin}, \cite{Ivanov RPJ 13}, \cite{Ivanov arxiv big 14} for the current-current interaction. The third terms describes action of the external magnetic field on magnetic moments and the spin-spin interaction. The spin-spin interaction is the dipole-dipole interaction of the magnetic moments via the magnetic field created by the magnetic moments. Generally speaking, the spin-current interaction gives contribution in this term. It comes via the magnetic field created by the electric currents and acting on the magnetic moments. We do not consider the spin-current interaction in this paper explicitly, for details see \cite{Andreev RPJ 07}. The fourth term in the right-hand side of equation (\ref{E-P_QP Euler equation for electrons}) (equation (\ref{E-P_QP Euler equation for positrons})) is the Darwin interaction between electrons (positrons) (see Ref. \cite{Landau 4} section 83 formula 15). The fifth terms in equations (\ref{E-P_QP Euler equation for electrons}) and (\ref{E-P_QP Euler equation for positrons}) are also the Darwin interaction between particles of different species. The last two terms in both equations are the annihilation interaction (see Ref. \cite{Pirenne 1947} and Ref. \cite{Landau 4} formula (83.24)). We see that the fifth and sixth terms are similar, but they have different nature, one of them is the Darwin interaction, another one is the the annihilation interaction. Hence we present them separately to make stress on their different nature.

We need to get closed set of equations, so we should use an equation of state for the pressure $p$. We consider degenerate electrons. Hence, in non-relativistic case, we have $p=p_{Fe}=\frac{(3\pi^{2})^{2/3}}{5}\frac{\hbar^{2}}{m}n^{5/3}$ \cite{Landau v5} (see section 57 formula 7). From this equation of state we find $\frac{\partial p}{\partial n}=\frac{1}{3}mv_{Fe}^{2}$, where $v_{Fe}=(3\pi^{2}n_{0})^{1/3}\hbar/m$ is the Fermi velocity.

\begin{figure}
\includegraphics[width=8cm,angle=0]{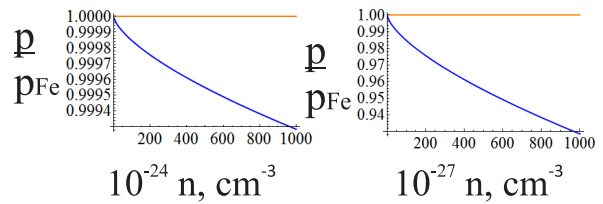}
\caption{\label{FD_Pressure} (Color online) The figure shows
decreasing of the pressure of degenerate electrons due to
semi-relativistic effects.}
\end{figure}

The spin-spin interaction, as well as the Darwin and the annihilation interactions, have semi-relativistic nature. Hence we should include semi-relativistic effects in the equation of state
\begin{equation}\label{E-P_QP rel eq of state Semi-rel} p=p_{Fe}\biggl(1-\frac{1}{14}\frac{v_{Fe}^{2}}{c^{2}}\biggr),\end{equation}
where $p_{Fe}$ and $v_{Fe}$ are the non-relativistic Fermi
pressure and Fermi velocity presented above. This formula shows
decreasing of pressure at large concentration caused by the
semi-relativistic effects. This decreasing is presented on Fig.
(\ref{FD_Pressure}).

Perturbation of pressure in the semi-relativistic  approach appears as
\begin{equation}\label{E-P_QP rel eq of state Semi-rel perturbation} \delta p=\frac{\partial p}{\partial n}\delta n=\frac{1}{3}mv_{Fe}^{2}\biggl(1-\frac{1}{10}\frac{v_{Fe}^{2}}{c^{2}}\biggr).\end{equation}

The Fermi velocity $v_{Fe}$ is determined by the equilibrium concentration $n_{0}$. The semi-relativistic part of the Fermi pressure is noticeable $\frac{v_{Fe}^{2}}{10c^{2}}\sim10^{-3}\div0.1$. It corresponds to $n_{0}\sim10^{27}$cm$^{-3}$.

\subsection{Magnetic moment (spin) evolution}

Explicit form of the magnetization (density of the magnetic moments proportional to the spin density) definition
\begin{equation}\label{E-P_QP def of magnetization} \textbf{M}_{a}(\textbf{r},t)=\int dR\sum_{i}\delta(\textbf{r}-\textbf{r}_{i})\gamma_{i}\psi^{+}(R,t)\mbox{\boldmath $\sigma$}_{i}\psi(R,t),\end{equation}
and the Schrodinger equation (\ref{E-P_QP Schrodinger eq}) allows to derive magnetic moment (spin) evolution equations for each species.
For electrons we find
$$n_{e}(\partial_{t}+\textbf{v}_{e}\nabla) \mbox{\boldmath $\mu$} _{e} -\frac{\hbar}{2m\gamma_{e}}\partial^{\beta}[n_{e} \mbox{\boldmath $\mu$}_{e} , \partial^{\beta}\mbox{\boldmath $\mu$} _{e} ]$$
\begin{equation}\label{E-P_QP eq of magnetic moments evol electr} =\frac{2\gamma_{e}}{\hbar}n_{e}[\mbox{\boldmath $\mu$}_{e},\textbf{B}]+2\pi \frac{2\gamma_{e}}{\hbar} n_{e}n_{p}[\mbox{\boldmath $\mu$}_{e}, \mbox{\boldmath $\mu$}_{p}],\end{equation}
and for positrons we obtained
$$n_{p}(\partial_{t}+\textbf{v}_{p}\nabla) \mbox{\boldmath $\mu$}_{p}-\frac{\hbar}{2m\gamma_{p}}\partial^{\beta}[n_{p}
\mbox{\boldmath $\mu$}_{p},\partial^{\beta}\mbox{\boldmath $\mu$}_{p}]$$
\begin{equation}\label{E-P_QP eq of magnetic moments evol positr} =\frac{2\gamma_{p}}{\hbar}n_{p}[\mbox{\boldmath $\mu$}_{p},\textbf{B}]+2\pi \frac{2\gamma_{p}}{\hbar} n_{e}n_{p}[\mbox{\boldmath $\mu$}_{p}, \mbox{\boldmath $\mu$}_{e}].\end{equation}
These equations are a generalization of the Bloch equation. The first groups of terms in the left-hand side of equations (\ref{E-P_QP eq of magnetic moments evol electr}) and (\ref{E-P_QP eq of magnetic moments evol positr}) are the substantial derivative of the reduced magnetic moments. The second terms are the quantum Bohm potential for the Bloch equation. In the right-hand side of equations (\ref{E-P_QP eq of magnetic moments evol electr}) and (\ref{E-P_QP eq of magnetic moments evol positr}) are the torque caused by the interaction with the external magnetic field and the interparticle interactions. The first terms in the right-hand side describes interaction of magnetic moments with the external magnetic field, with internal magnetic fields caused by the magnetic moments and the electric currents of the electrons and positrons. Interaction of magnetic moments with the electric currents is an example of the spin-current interaction, which is not included in this paper, but it can be included by the many-particle quantum hydrodynamic method \cite{Andreev RPJ 07}.

The quantum Bohm potential contribution in the magnetic moment balance equation is written in the single particle approximation. This is the same approximation, which is used for the quantum Bohm potential in the Euler equations. Explanation of this approach is presented in the previous subsection.

Usually, dealing with several species of particles one solves hydrodynamic equation for each species and put the result in the Maxwell equations. However, if we consider electron-positron plasmas with the annihilation interaction we find that hydrodynamic variables of different species are mixed. We see that the Euler equation for electrons (positrons) contains the concentration $n_{p}$ ($n_{e}$) and the reduced magnetization $\mbox{\boldmath $\mu$}_{p}$ ($\mbox{\boldmath $\mu$}_{p}$) of positrons (electrons). Similar situation we have with the magnetic moment evolution equation, where $\partial_{t}\mbox{\boldmath $\mu$}_{e}$ ($\partial_{t}\mbox{\boldmath $\mu$}_{p}$) contains $n_{p}$ and $\mbox{\boldmath $\mu$}_{p}$ ($n_{e}$ and $\mbox{\boldmath $\mu$}_{e}$). Hence we should solve altogether the hydrodynamic equations for both species. As the result we find $n_{e}$, $n_{p}$, $\textbf{v}_{e}$, $\textbf{v}_{p}$, $\mbox{\boldmath $\mu$}_{e}$, and $\mbox{\boldmath $\mu$}_{p}$ as functions of electric $\textbf{E}$ and magnetic $\textbf{B}$ fields. So, we can put our results in the Maxwell equations.

One more method of QHD derivation has been suggested recently \cite{Koide PRC 13}. This method also gives full chain of hydrodynamic equations. Particularly, in Ref. \cite{Koide PRC 13} author discuss derivation the energy density, the positivity of the entropy production
is used in the derivation

\subsection{Maxwell equations}

Electromagnetic fields of interaction between particles coupled with their sources by means the Maxwell equation
\begin{equation}\label{E-P_QP div E} \nabla \textbf{E}=4\pi\sum_{a}q_{a}n_{a},\end{equation}
\begin{equation}\label{E-P_QP div B} \nabla \textbf{B}=0, \end{equation}
\begin{equation}\label{E-P_QP ror E} \nabla\times \textbf{E}=-\frac{1}{c}\partial_{t}\textbf{B},\end{equation}
and
\begin{equation}\label{E-P_QP rot B}\nabla\times \textbf{B}=\frac{1}{c}\partial_{t}\textbf{E}
+\frac{4\pi}{c}\sum_{a}q_{a}n_{a}\textbf{v}_{a}+4\pi\sum_{a}\nabla\times (n_{a}\mbox{\boldmath $\mu$}_{a}).\end{equation}

In our theory we apply the many-particle Schrodinger equation, which is a non-relativistic equation. Hence we do not obtain time derivatives in the Maxwell equations. Our model is an analog of the Coulomb plasmas. We have quasi-static Coulomb and spin-spin interactions leading to quasi-static Maxwell equations \cite{Andreev RPJ 07}. Nevertheless we have generalized obtained equations to the full set of Maxwell equations.

Equations of quantum hydrodynamics derived in this paper are obtained in the self-consistent field approximation. They do not include exchange interaction. QHD including exchange interaction were developed in Refs. \cite{MaksimovTMP 1999}, \cite{MaksimovTMP 2001 b}, \cite{Andreev 1403 exchange}. Exchange part of the Coulomb interaction was derived in Ref. \cite{MaksimovTMP 1999} for bosons and fermions. Force field of the exchange  Coulomb and spin-spin interactions between spin-1/2 fermions was derived in Ref. \cite{MaksimovTMP 2001 b}. More general approach for the Coulomb exchange interaction of fermions in three- and two- dimensional quantum plasmas was developed in Ref. \cite{Andreev 1403 exchange}. It was applied to spectrum of collective excitations in Refs. \cite{Andreev AtPhys 08}, \cite{Andreev 1403 exchange}.

The QHD model presented by formulae (\ref{E-P_QP continuity eq for electron})-(\ref{E-P_QP rot B}) is obtained for point-like particles. Classic and quantum hydrodynamics for finite size ions and dust particles was developed in Ref. \cite{Andreev 1401 finite ions}. We do not need to include finite size of particles in theory of electron-positron plasmas, since modern fundamental theory consider these particles as point-like objects.

\section{Dispersion of waves propagating parallel to the external magnetic field}

We consider propagation of small perturbations around an equilibrium. We assume that plasma is in a constant uniform external magnetic field. Hence we have
$$\begin{array}{ccc} n_{a}=n_{0a}+\delta n_{a}, &\textbf{E}=0+\delta\textbf{E},\end{array}$$
$$\begin{array}{ccc} \textbf{B}=B_{0}\textbf{e}_{z}+\delta\textbf{B},&
\textbf{v}_{a}=0+\textbf{v}_{a},& \mbox{\boldmath $\mu$}_{a}=\mbox{\boldmath $\mu$}_{0a}+\delta\mbox{\boldmath $\mu$}_{a},\end{array}$$
\begin{equation}\label{E-P_QP}\begin{array}{ccc} &
\textbf{M}_{0a}=n_{0a}\mbox{\boldmath $\mu$}_{0a}=\chi_{a}\textbf{B}_{0}, & n_{0e}=n_{0p}. \end{array}\end{equation}

Electrons and positrons are polarized by the external magnetic field, so their reduced magnetizations $\mbox{\boldmath $\mu$}_{0a}$ equal to each other
\begin{equation}\label{E-P_QP} \mbox{\boldmath $\mu$}_{0e}=\mbox{\boldmath $\mu$}_{0p}. \end{equation}
Thus they are directed in direction of the external field. Electrons and positrons have different electric charge $\pm e$. Consequently signs of the gyromagnetic ratio are also different. Hence equilibrium spins of electrons and ions have different direction.

Dielectric tensor for spin-1/2 magnetized electron-positron plasmas has diagonal form with $\varepsilon_{xx}=\varepsilon_{yy}\neq0$ and $\varepsilon_{zz}\neq0$. Other elements are equal to zero $\varepsilon_{xy}=\varepsilon_{xz}=\varepsilon_{yx}=\varepsilon_{yz}=\varepsilon_{zx}=\varepsilon_{zy}=0$.

\subsection{Longitudinal waves}

General form of dispersion equation for the longitudinal waves in magnetized electron-positron plasmas at presence of the annihilation interaction appears as
\begin{equation}\label{E-P_QP} -1-\frac{2\omega_{Le}^{2}}{\Gamma^{2}-k^{2}\Xi^{2}}=0, \end{equation}
where
\begin{equation}\label{E-P_QP} \omega_{Le}^{2}=\frac{4\pi e^{2}n_{0}}{m}\end{equation}
is the Langmuir frequency,
$$\Gamma^{2}\equiv-\omega^{2}+\frac{1}{3}v_{Fe}^{2}\biggl(1-\frac{1}{10}\frac{v_{Fe}^{2}}{c^{2}}\biggr)k_{z}^{2}$$
\begin{equation}\label{E-P_QP}
-\frac{\pi e^{2}\hbar^{2}}{m^{3}c^{2}}n_{0}k_{z}^{2}+\frac{\hbar^{2}k_{z}^{4}}{4m^{2}},\end{equation}
and
\begin{equation}\label{E-P_QP Xi def}  \Xi^{2}=n_{0}\biggl(\frac{5}{2}\frac{\pi e^{2}\hbar^{2}}{m^{3}c^{2}}-\frac{2\pi}{m}\mu_{0e}\mu_{0p}\biggr).\end{equation}
Since $\mu_{0,max}=\frac{e\hbar}{2mc}$ and $\mu_{0}<\sqrt{5}\frac{e\hbar}{2mc}$ for electrons and positrons, when $\Xi^{2}>0$.

In absence of the annihilation interaction we have $\Xi^{2}=n_{0}\frac{\pi e^{2}\hbar^{2}}{m^{3}c^{2}}$ existing due to the electron-positron Darwin interaction, when we obtain
the well-known Langmuir wave
$$\omega^{2}=2\omega_{Le}^{2}-2\frac{\pi e^{2}\hbar^{2}}{m^{3}c^{2}}n_{0}k_{z}^{2}$$
\begin{equation}\label{E-P_QP DE longit simplified Lang} +\frac{1}{3}v_{Fe}^{2}\biggl(1-\frac{1}{10}\frac{v_{Fe}^{2}}{c^{2}}\biggr)k_{z}^{2}+\frac{\hbar^{2}k_{z}^{4}}{4m^{2}}, \end{equation}
containing contribution of the Coulomb interaction, the Darwin interaction, the Fermi pressure, and the quantum Bohm potential, correspondingly. For electron-ion plasmas one does not find coefficient 2 before the second term. The contribution of the Darwin interaction is doubling in electron-positron plasmas in compare with the electron-ion plasmas. It happens due to equality of masses of electrons and positrons.

Existence of the annihilation interaction leads to an extra term modifying the Langmuir wave spectrum. We present dispersion of the longitudinal waves as
$$\omega^{2}=2\omega_{Le}^{2}-2\frac{\pi e^{2}\hbar^{2}}{m^{3}c^{2}}n_{0}k_{z}^{2}+\frac{1}{3}v_{Fe}^{2}\biggl(1-\frac{1}{10}\frac{v_{Fe}^{2}}{c^{2}}\biggr)k_{z}^{2}$$
\begin{equation}\label{E-P_QP DE longit general Lang} +\frac{\hbar^{2}k_{z}^{4}}{4m^{2}}-n_{0}\biggl(\frac{3}{2}\frac{\pi e^{2}\hbar^{2}}{m^{3}c^{2}}-\frac{2\pi}{m}\mu_{0}^{2}\biggr)k_{z}^{2}, \end{equation}
where the last term describes the contribution of the annihilation interaction. In formula (\ref{E-P_QP DE longit general Lang}) we separate two parts of $\Xi^{2}$ appearing from different interactions.

Let us rewrite formula in different form getting together similar terms coming from the Darwin interaction and the spinless part of annihilation interaction
$$\omega^{2}=2\omega_{Le}^{2}\biggl(1-\frac{7}{16}\frac{\hbar^{2}}{m^{2}c^{2}}k_{z}^{2}\biggr)$$
\begin{equation}\label{E-P_QP DE longit general Lang 2} +\frac{1}{3}v_{Fe}^{2}\biggl(1-\frac{1}{10}\frac{v_{Fe}^{2}}{c^{2}}\biggr)k_{z}^{2}+\frac{\hbar^{2}k_{z}^{4}}{4m^{2}}+\frac{2\pi n_{0}}{m}\mu_{0}^{2}k_{z}^{2}. \end{equation}

\begin{figure}
\includegraphics[width=8cm,angle=0]{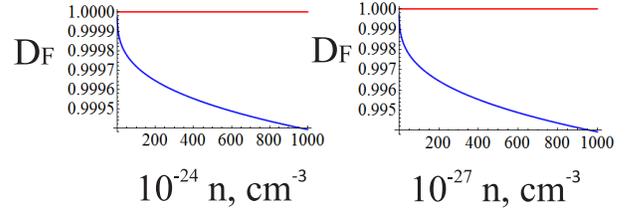}
\caption{\label{FD_FermiVsDarwin} (Color online) The figure shows
comparison of the non-relativistic part of Fermi pressure with the
simultaneous contribution of the Darwin and annihilation
interaction appearing in the Langmuir wave spectrum.}
\end{figure}

\begin{figure}
\includegraphics[width=8cm,angle=0]{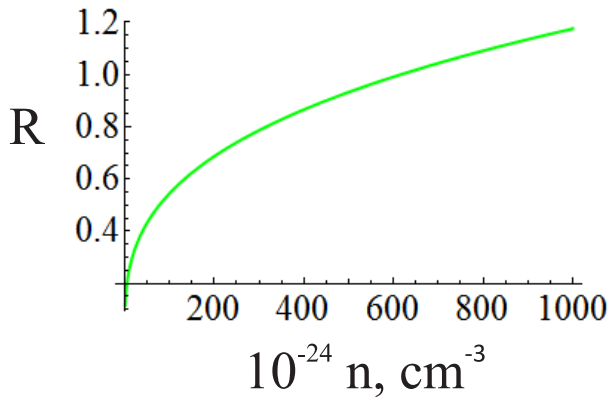}
\caption{\label{FD_RatioOfFermiShiftToDarwin} (Color online) The figure shows
ratio of the semi-relativistic part of Fermi pressure to the
simultaneous contribution of the Darwin and annihilation
interaction contributions $R=\Delta\omega_{RF}/\Delta\omega_{AD}$ appearing in the Langmuir wave spectrum, where $\Delta\omega_{RF}=-\frac{1}{30}\frac{v_{Fe}^{4}}{c^{2}}k_{z}^{2}$, $\Delta\omega_{AD}=-\frac{7}{2}\frac{\pi e^{2}\hbar^{2}}{m^{3}c^{2}}n_{0}k_{z}^{2}$.}
\end{figure}

Simultaneous action of the Darwin and annihilation interactions
gives
$A\equiv\frac{n_{0}}{m}(2\pi\mu_{0}^{2}-\frac{7}{2}\pi\frac{e^{2}\hbar^{2}}{m^{2}c^{2}})k_{z}^{2}$.
Assuming $\mu_{0}=\mu_{0,max}=\frac{e\hbar}{2mc}$ we have minimal
contribution of these interactions in the spectrum. Maximal
contribution of these interactions in the spectrum of Langmuir
waves we find at $\mu_{0}\approx0$. It can happen at small or zero
external magnetic fields. Comparing quantity $A$ with the
non-relativistic part of Fermi pressure at $\mu_{0}=\mu_{0,max}$
we can represent their difference as
$3n_{0}^{2/3}\frac{\hbar^{2}}{m^{2}}[(\frac{\pi}{3})^{4/3}-\pi\frac{e^{2}n_{0}^{1/3}}{mc^{2}}]$,
where the last term in the brackets, presenting $A$, is the ratio
of the average Coulomb interaction to the rest energy of electron.
The Darwin and annihilation interaction give contribution in the
Langmuir wave spectrum at large densities. At
$n_{0}=10^{27}$cm$^{-3}$ $A$ gives 0.05 percent of the Fermi
pressure. Increasing equilibrium concentration up to
$n_{0}=10^{30}$cm$^{-3}$ we find that $A$ is 0.5 percent of the
Fermi pressure. This comparison is summarized on Fig.
(\ref{FD_FermiVsDarwin}). Comparing figures (\ref{FD_Pressure}) and (\ref{FD_FermiVsDarwin}) we see that the Darwin and annihilation interactions give larger contribution than the semi-relativistic part of the pressure at particle concentrations lower than $10^{26}$ cm$^{-3}$. At particle concentrations larger $10^{26}$ cm$^{-3}$ the semi-relativistic part of the pressure dominates over the simultaneous action of the Darwin and annihilation interactions (see Fig. (\ref{FD_RatioOfFermiShiftToDarwin})).

\subsubsection{Darwin interaction contribution}

In nature the spin-spin, Darwin, and annihilation interactions are the semi-relativistic (weakly-relativistic) effects (see \cite{Landau 4} for instance). However there are more semi-relativistic effects. They are the spin-current \cite{Andreev RPJ 07} and the current-current \cite{Ivanov Darwin}, \cite{Ivanov arxiv big 14}, \cite{Ivanov RPJ 13} interactions, mentioned above, the relativistic correction to kinetic energy (RCKE) \cite{Ivanov Darwin}, \cite{Ivanov arxiv big 14}, \cite{Ivanov RPJ 13}, and the spin-orbit interaction as well \cite{pavelproc}, \cite{Andreev IJMP 12}. All these interactions are parts of the Breit Hamiltonian (see \cite{Landau 4} section 83). Strictly speaking all these effects should be considered simultaneously. Contributions of these interactions in hydrodynamic or kinetic equations for quantum plasmas reveal different behavior. Hence, sometimes, these interactions may be considered separately, as we do it now, considering the spin-spin, Darwin, and annihilation interactions only. But it is important to remember behavior of other interactions. We should mention that usually in literature authors consider the spin-spin interaction model. In this paper we show that, for plasmas with positrons, it is necessary to expand such model towards account of the annihilation interaction.

A change of the Darwin interaction contribution due to simultaneous account of the Coulomb interaction and the semi-relativistic amendment to the kinetic energy was found in Ref. \cite{Ivanov Darwin}. We do not consider the semi-relativistic part of kinetic energy in this paper. We present solutions, which directly follows from microscopic theory we have presented. Nevertheless we should keep in mind the result of Ref. \cite{Ivanov Darwin}.

In Ref. \cite{Ivanov Darwin}, the change of the electron-electron Darwin interaction is shown. In electron-positron plasmas, in the same way, we find the change of the electron-positron and positron-positron Darwin interactions as well.

The Hamiltonian of the RCKE, giving result (\ref{E-P_QP Delta omega C r}), appears as
\begin{equation}\label{E-P_QP}\hat{H}_{r}=-\sum_{i}\frac{\widehat{\textbf{D}}_{i}^{4}}{8m_{i}^{3}c^{2}}.\end{equation}
It is the second order contribution at expansion of the relativistic kinetic energy $E_{kin}=\sqrt{p^{2}c^{2}+m^{2}c^{4}}-mc^{2}$ in the series at the weakly-relativistic limit.

Simultaneous consideration of the RCKE and interaction of charges with the external electric field the interparticle Coulomb field gives the following force field in the Euler equation
\begin{equation}\label{E-P_QP}\textbf{F}_{r}=\frac{q\hbar^{2}}{4m^{2}c^{2}}\partial^{\beta}(n\nabla E^{\beta}).\end{equation}
Hence the contribution of the electric field in the Euler equation appears as the combination of non-relativistic and semi-relativistic parts
$\textbf{F}_{C,r}=q_{a}n_{a}\textbf{E}+\textbf{F}_{r}$.

It was found in Refs. \cite{Ivanov Darwin} and \cite{Ivanov RPJ 13} that for electron-ion
plasmas the shift of the frequency square due to the RCKE is
\begin{equation}\label{E-P_QP Delta omega C r}\Delta\omega_{C,r}=-\omega_{Le}^{2}\frac{\hbar^{2}k^{2}}{4m^{2}c^{2}}=-\frac{\pi e^{2}\hbar^{2}}{m^{3}c^{2}}n_{0}k^{2}.\end{equation}
For electron-positron plasmas we find
\begin{equation}\label{E-P_QP Delta omega C r e-p}\Delta\omega_{C,r}=-2\omega_{Le}^{2}\frac{\hbar^{2}k^{2}}{4m^{2}c^{2}}=-\frac{2\pi e^{2}\hbar^{2}}{m^{3}c^{2}}n_{0}k^{2}\end{equation}
instead of (\ref{E-P_QP Delta omega C r}).

Spectrum of the Langmuir waves propagating parallel to the external magnetic field with account of the RCKE is obtained as
$$\omega^{2}=2\omega_{Le}^{2}\Biggl(1-\biggl(\frac{3}{16}+\frac{1}{2}\biggr)\frac{\hbar^{2}}{m^{2}c^{2}}k_{z}^{2}\Biggr)+\frac{1}{3}v_{Fe}^{2}k_{z}^{2}$$
\begin{equation}\label{E-P_QP DE longit general Lang 2} +\frac{\hbar^{2}k_{z}^{4}}{4m^{2}}+\frac{2\pi n_{0}}{m}\mu_{0}^{2}k_{z}^{2}+ThRE, \end{equation}
where $ThRE$ means contribution of thermal-relativistic effects.
The sum $(\frac{3}{16}+\frac{1}{2})$ depicts sum of contribution
of the annihilation interaction (3/16) and combination of the
Darwin interaction with the semi-relativistic correction of the
Coulomb interaction.

\subsection{Transverse waves}

In absence of the annihilation interaction, for the transverse waves, we find the dispersion equation
\begin{equation}\label{E-P_QP} k_{z}^{2}c^{2}-\omega^{2}+2\omega_{Le}^{2} \frac{\omega^{2}}{\omega^{2}-\Omega^{2}}+\frac{2\gamma}{\hbar}\frac{8\pi(k_{z}c)^{2}n_{0}\mu_{0}\Omega_{\gamma}}{\omega^{2}-\Omega_{\gamma}^{2}}=0, \end{equation}
with $\Omega=\frac{eB_{0}}{mc}$ is the cyclotron frequency for a charge in magnetic field, $\Omega_{\gamma}=\frac{2\gamma}{\hbar}B_{0}+\frac{\hbar}{2m\gamma}\mu_{0}k_{z}^{2}$ is the shifted cyclotron frequency for a magnetic moment in external magnetic field, it is shifted due to the quantum Bohm potential contribution in the magnetic moment evolution equation.

Equation is an analog of dispersion equation for spin-1/2 electron-ion plasmas \cite{Misra JPP 10}, \cite{Andreev IJMP 12}, \cite{pavelproc}. In such case circular polarized waves propagate in electron-ion plasmas. But electron-positron plasmas we have linearly polarized wave as it is in spinless classic magnetized plasmas.

In presence of the annihilation interaction the dispersion equation for the transverse waves appears as
$$k_{z}^{2}c^{2}-\omega^{2}+2\omega_{Le}^{2} \frac{\omega^{2}}{\omega^{2}-\Omega^{2}}$$
\begin{equation}\label{E-P_QP transverse wave with annihil} +\frac{2\gamma}{\hbar}\frac{8\pi(k_{z}c)^{2}n_{0}\mu_{0}\biggl(\Theta+\frac{2\gamma}{\hbar}2\pi n_{0}\mu_{0}\biggr)}{\omega^{2}-\Theta^{2}+\biggl(\frac{2\gamma}{\hbar}2\pi n_{0}\mu_{0}\biggr)^{2}}=0, \end{equation}
where
\begin{equation}\label{E-P_QP Theta def} \Theta=\frac{2\gamma}{\hbar}B_{0}+\frac{\hbar}{2m\gamma}\mu_{0}k_{z}^{2}+\frac{2\gamma}{\hbar}2\pi n_{0}\mu_{0}.\end{equation}
$\Theta$ contains and extra term in compare with $\Omega_{\gamma}$. Thus $\Theta$ is is twice shifted cyclotron frequency of the magnetic moment in the external magnetic field $\frac{2\gamma}{\hbar}B_{0}$. New shift is presented by the last term in formula (\ref{E-P_QP Theta def}). This term is caused by the annihilation interaction.

Papers \cite{Misra JPP 10}, \cite{Andreev IJMP 12}, \cite{pavelproc} did not included contribution of the quantum Bohm potential in the spin current in the magnetic moment evolution equation. In this paper the quantum Bohm potential is included. In dispersion equation for transverse waves, the quantum Bohm potential is presented by the second term in the right-hand side of formula (\ref{E-P_QP transverse wave with annihil}).

In absence of magnetic field we have
\begin{equation}\label{E-P_QP Disp of TR waves no MF} \omega^{2}=2\omega_{Le}^{2}+k_{z}^{2}c^{2}.\end{equation}
Presence of the magnetic field with no spin account gives
$$\omega^{2}_{01,02}=\omega_{Le}^{2}+\frac{1}{2}k_{z}^{2}c^{2}+\frac{1}{2}\Omega^{2}$$
\begin{equation}\label{E-P_QP Disp of TR waves with MF}  \pm\sqrt{\biggl(\omega_{Le}^{2}+\frac{1}{2}k_{z}^{2}c^{2}+\frac{1}{2}\Omega^{2}\biggr)^{2}-k_{z}^{2}c^{2}\Omega^{2}}.\end{equation}

Solution with plus before the square root in formula (\ref{E-P_QP Disp of TR waves with MF}) gives a solution with $\omega^{2}$ lower than $2\omega_{Le}^{2}+k^{2}c^{2}+\Omega^{2}$ due to the last term under the square root $-k_{z}^{2}c^{2}\Omega^{2}$.

In the case of minus before the square root in formula (\ref{E-P_QP Disp of TR waves with MF}) we can rewrite this formula as
\begin{equation}\label{E-P_QP Disp of TR waves with MF with minus separately} \omega^{2}=\frac{k_{z}^{2}c^{2}\Omega^{2}}{\Upsilon_{z} +\sqrt{\Upsilon_{z}^{2}-k_{z}^{2}c^{2}\Omega^{2}}},\end{equation}
where $\Upsilon_{z}=\omega_{Le}^{2}+\frac{1}{2}k_{z}^{2}c^{2}+\frac{1}{2}\Omega^{2}$.

Next, account of the magnetic moment along with the annihilation interaction gives shift of solutions (\ref{E-P_QP Disp of TR waves with MF}), where we assumed that contribution of spin is small far from resonance frequency of the last term in formula (\ref{E-P_QP transverse wave with annihil}), when it denominator close to zero.

Shifts of frequencies (\ref{E-P_QP Disp of TR waves with MF}) due to spin of particles appears as
$$\delta\omega_{i}=\frac{2\gamma}{\hbar}\frac{8\pi(k_{z}c)^{2}n_{0}\mu_{0}\biggl(\Theta+\frac{2\gamma}{\hbar}2\pi n_{0}\mu_{0}\biggr)}{\omega_{0i}^{2}-\Theta^{2}+\biggl(\frac{2\gamma}{\hbar}2\pi n_{0}\mu_{0}\biggr)^{2}}\times$$
\begin{equation}\label{E-P_QP} \times\frac{\omega_{0i}^{2}-\Omega^{2}}{2\omega_{0i}(2\omega_{0i}^{2}-2\omega_{Le}^{2}-k_{z}^{2}c^{2}-\Omega^{2})}\end{equation}
Hence full solutions look like $\omega_{1,2}=\omega_{01,02}+\delta\omega_{1,2}$.

\subsubsection{Spin-plasma waves}

Frequency of the spin-plasma waves is close to the resonance frequency of the last term in formula (\ref{E-P_QP transverse wave with annihil}). The resonance frequency has form of $\omega_{Res}^{2}=\Theta^{2}-\Lambda^{2}$, with $\Lambda=\frac{2\gamma}{\hbar}2\pi n_{0}\mu_{0}$.
We are looking for solution with frequencies close to the resonance frequency $\omega=\omega_{Res}+\delta\omega_{s}$.
In this case we have two regimes. They are regimes of small and "large" difference of two parameters. One of the two parameters $\delta\omega_{s}$ is difference of the spin-plasma wave frequencies from the resonance frequency $\omega_{Res}$.
The second parameter is the defect of frequency $\Delta^{2}=\Theta^{2}-\Lambda^{2}-\Omega^{2}$ showing difference of the resonant frequency $\omega_{Res}=\sqrt{\Theta^{2}-\Lambda^{2}}$ of the last term in equation (\ref{E-P_QP transverse wave with annihil}) from the electron the charge cyclotron frequency $\Omega=eB_{0}/(mc)$.

\begin{figure}
\includegraphics[width=8cm,angle=0]{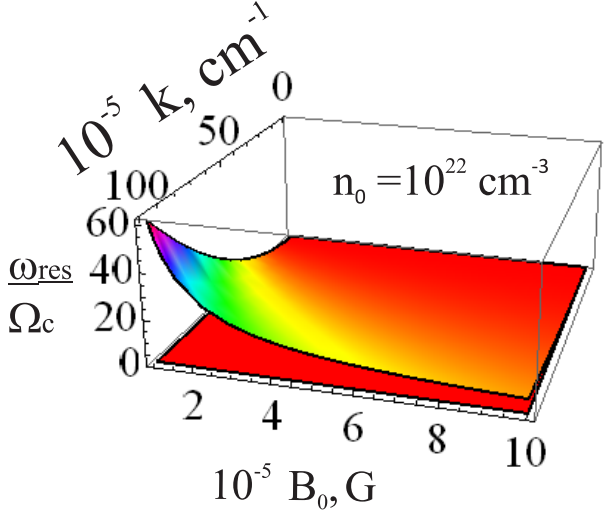}
\caption{\label{FD_SW1} (Color online) The figure shows dependence of the resonance frequency on the wave vector at different external magnetic fields. The figure is made for regime of large magnetization. We see large difference between the resonance frequency and the cyclotron frequency. This difference appear due to the annihilation interaction affecting the spin-plasma wave in magnetized electron-positron plasmas.}
\end{figure}

\begin{figure}
\includegraphics[width=8cm,angle=0]{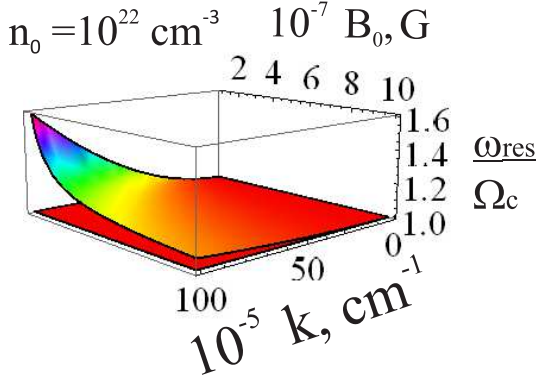}
\caption{\label{FD_SW2} (Color online) The figure shows dependence of the resonance frequency on the wave vector at different external magnetic fields in regime of rather small magnetization in compare with Fig. (\ref{FD_SW1}). Nevertheless the magnetization contribution is large enough to distinguish the resonance frequency and the cyclotron frequency.}
\end{figure}

Explicit form of the resonance frequency $\omega_{Res}$ at small magnetization $n_{0}\mu_{0}$ in compare with the external field $B_{0}$ appears as
\begin{equation}\label{E-P_QP omega Res small M} \omega_{Res}=\frac{2\gamma}{\hbar}B_{0}+\frac{\hbar}{2m\gamma}\mu_{0}k_{z}^{2}+\frac{2\gamma}{\hbar}2\pi n_{0}\mu_{0},\end{equation}
where different terms have the following physical meaning. The first term is the cyclotron frequency describing frequency of rotation of the magnetic moment in external magnetic field. The second term is the contribution of the quantum Bohm potential existing in the magnetic moment evolution equation. The last term presents the spin dependent part of the annihilation interaction, which appears in the magnetic moment evolution equations.

Assuming full polarization of spins we have $\mu_{0}=\gamma$. Using $\frac{2\gamma}{\hbar}B_{0}=g\Omega_{c}$ with $g=1.00116$ we can represent formula (\ref{E-P_QP omega Res small M}) as
\begin{equation}\label{E-P_QP omega Res small M via Omega c} \omega_{Res}=g\Omega_{c}+\frac{\hbar}{2m\gamma}\mu_{0}k_{z}^{2}+\frac{4\pi n_{0}g^{2}\gamma^{2}}{\hbar} .\end{equation}

If the equilibrium magnetization $M_{0}=n_{0}\mu_{0}$ is comparable with the external magnetic field, or it is much more than the external magnetic field we have
\begin{equation}\label{E-P_QP omega Res large M} \omega_{Res}=g\Omega_{c} \biggl(1+\frac{\hbar k_{z}^{2}}{2mg\Omega_{c}}\biggr)\sqrt{1+4\pi\frac{M_{0}}{B_{0}}}.\end{equation}
This formula corresponds to the long-wavelength limit, so we assumed that the quantum Bohm potential gives small contribution. When the contribution of magnetization dominates we can rewrite formula (\ref{E-P_QP omega Res large M}) as $\omega_{Res}=2\sqrt{\pi}\frac{e\sqrt{M_{0}B_{0}}}{mc}$.

At small magnetization in compare with the external magnetic field the annihilation interaction gives small shift of the resonance frequency from the cyclotron frequency of electrons. However it gives considerable contribution at large magnetization. The quantum Bohm potential reveals in short-wavelength limit, when $k$ is comparable with $\sqrt[3]{n_{0}}$. Magnetization $M_{0}$ depend on the degree of spin polarization and particles concentration $M_{0}=\mu_{0}n_{0}$. We consider large enough particle concentration and magnetic field to rich full spin polarization exceeding the external magnetic field.

At the first step we can take $n_{0}=10^{22}$ cm$^{-3}$ and $B_{0}=10^{4}$ G. Considering the long-wavelength limit we neglect the quantum Bohm potential. We find $\frac{\omega_{Res}}{\Omega_{c}}=0.063g\approx0.063$. In this case the annihilation interaction exceeds the anomalous magnetic moment of ten times. The short-wavelength limit for different magnetic fields is presented on Figs. (\ref{FD_SW1}) and (\ref{FD_SW2}).

\begin{figure}
\includegraphics[width=8cm,angle=0]{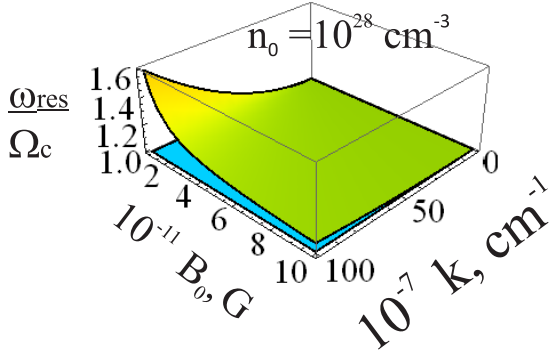}
\caption{\label{FD_SW3} (Color online) The figure shows dependence of the resonance frequency on the wave vector at different external magnetic fields in regime of large particle concentration and large magnetic field. Large concentration allows to reach large magnetization and rather small wavelengths. Increasing of the external magnetic field increases the cyclotron frequency and the resonance frequency. The figure presents relative shift caused by the quantum Bohm potential and the annihilation interaction.}
\end{figure}

Next let us make step towards larger concentrations and magnetic fields. Large concentrations open possibilities to reach large wave vectors and large magnetization. Simultaneous increasing of the concentration and magnetic field gives a behavior depicted on Fig. (\ref{FD_SW3}), which is  similar to considered above and presented on Fig. (\ref{FD_SW2}).

Increasing of the wave vector or magnetization without change of the magnetic field increasing the resonance frequency with simultaneous increasing of difference between the resonance frequency and cyclotron frequency.

\begin{figure}
\includegraphics[width=8cm,angle=0]{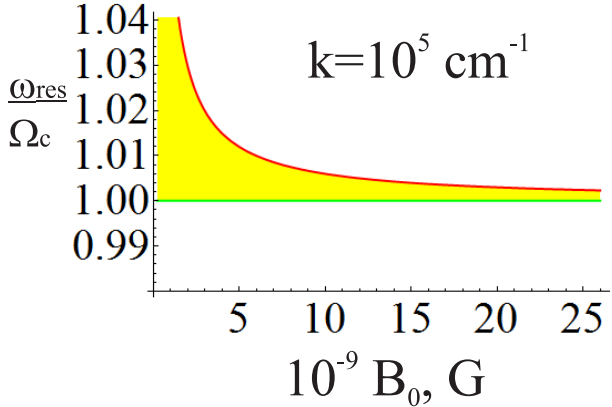}
\caption{\label{FD_SW4} (Color online) The figure shows behavior of the resonance frequency at fixed wave vector and different external magnetic fields. Horizontal line presents the unshifted cyclotron frequency to compare with the resonance frequency. Color filling between lines shows increasing of the resonance frequency due to the quantum Bohm potential and annihilation interaction.}
\end{figure}

Change of the resonance frequency with increasing of the external magnetic field at fixed wave vector and particle concentration is shown on Fig. (\ref{FD_SW4}).

\begin{figure}
\includegraphics[width=8cm,angle=0]{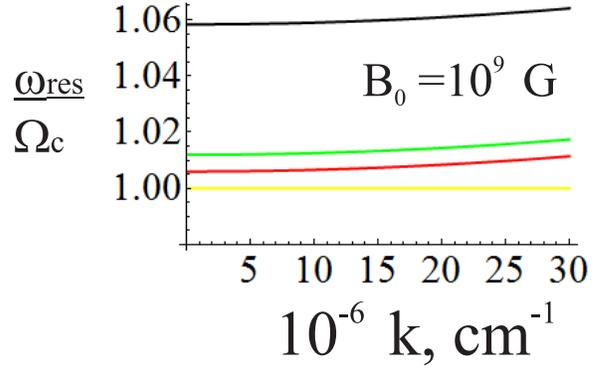}
\caption{\label{FD_SW5} (Color online) The figure shows dispersion dependence for the resonance frequency determining dispersion dependence of the spin-plasma wave in magnetized electron-positron plasmas. Dependence of the resonance frequency on the wave vector exists in electron-ion plasmas as well. On this picture we presents $\omega_{Res}(k)$ at fixed magnetic field. We consider the dispersion dependence in fixed interval of wave vectors at different particle concentration and, consequently, at different magnetization. This dependence does not exist in the electron-ion plasmas since it caused by the annihilation interaction. Increasing of the magnetization increases the resonance frequency. Consequently it increases difference between the the resonance frequency and the cyclotron frequency. Horizontal line at $\omega_{Res}/\Omega_{c}$ presents the cyclotron frequency (yellow line). Upper curves show the resonance frequency. The resonance frequencies increases with increasing of the concentration. We chouse the following values of the equilibrium concentration: $n_{0}=10^{27}$ cm$^{-3}$ (red curve), $n_{0}=2\cdot10^{27}$ cm$^{-3}$ (green curve), $n_{0}=10\cdot10^{27}$ cm$^{-3}$ (black curve). }
\end{figure}

Increasing the resonance frequency with increasing of the particle concentration due to the annihilation interaction is presented on Fig. (\ref{FD_SW5}).

In the limit of large frequency shift $\delta\omega$ in compare with $\Delta$, when
$\Delta^{2}\ll\delta\omega\sqrt{\Theta^{2}-\Lambda^{2}}$, we find
$$\delta\omega_{s}=-\frac{\Theta+\Lambda}{\sqrt{\Theta^{2}-\Lambda^{2}}} \times$$
\begin{equation}\label{E-P_QP spin wave paral shift large} \times\frac{\omega_{Le}^{2}(\Theta-\Lambda)+4\pi\frac{2\gamma}{\hbar}(k_{z}c)^{2}n_{0}\mu_{0}}{k^{2}c^{2}-\Theta^{2}+\Lambda^{2}}. \end{equation}

Let us represent formula (\ref{E-P_QP spin wave paral shift large}) in more explicit form in approximation of small magnetization and long wavelength
\begin{equation}\label{E-P_QP spin wave parall small things} \omega_{s}=\frac{2\gamma}{\hbar}B_{0}\biggl(1+\frac{2\gamma}{\hbar}\frac{4\pi n_{0}\mu_{0}}{B_{0}}\biggr)\frac{\omega_{Le}^{2}}{k^{2}c^{2}-(\frac{2\gamma}{\hbar}B_{0})^{2}}. \end{equation}

In opposite limit, for small frequency shift $\delta\omega$, at
$\Delta^{2}\gg\delta\omega\sqrt{\Theta^{2}-\Lambda^{2}}$, we obtain
$$\delta\omega_{s}=-\frac{2\gamma}{\hbar}\frac{\Theta+\Lambda}{\sqrt{\Theta^{2}-\Lambda^{2}}}$$
\begin{equation}\label{E-P_QP} \times\frac{4\pi(k_{z}c)^{2}n_{0}\mu_{0}}{\biggl(k_{z}^{2}c^{2}-\Theta^{2}+\Lambda^{2}+2\omega_{Le}^{2}\frac{(\Theta^{2}-\Lambda^{2})}{\Delta^{2}}\biggr)}. \end{equation}
Similar solution for electron-ion plasmas were found in Ref. \cite{Misra JPP 10}, \cite{pavelproc} (see formulas 2 and 3, for electrons and ions correspondingly), \cite{Andreev IJMP 12} (see formulas 10 and 11).

\section{Dispersion of waves propagating perpendicular to the external magnetic field}

From equation for $\delta E_{x}$ we find dispersion of longitudinal waves
$$\omega^{2}=2\omega_{Le}^{2}+\Omega^{2}+\frac{1}{3}v_{Fe}^{2}\biggl(1-\frac{1}{10}\frac{v_{Fe}^{2}}{c^{2}}\biggr)k_{x}^{2}$$
\begin{equation}\label{E-P_QP Langmuir wave perpendicular} -\frac{7}{2}\frac{\pi e^{2}\hbar^{2}}{m^{3}c^{2}}n_{0}k_{x}^{2} +\frac{2\pi n_{0}}{m}\mu_{0}^{2}k_{x}^{2}+\frac{\hbar^{2}k_{x}^{4}}{4m^{2}}.\end{equation}
This formula has simple structure. The first term is the Langmuir frequency caused by the Coulomb interaction. The second term is related to motion of a charge in an external magnetic field. The third term is the contribution of the Darwin interaction. Let us point out the remark we have made in previous section about the Darwin interaction contribution. The fourth and sixth terms are the contribution of the Fermi pressure and the quantum Bohm potential correspondingly. The last term is caused by the annihilation interaction.

Transverse waves describing by equation of $\delta E_{y}$ evolution, in absence of magnetic moment of particles, have the following dispersion equation
$$(\omega^{2}-k_{x}^{2}c^{2}-2\omega_{Le}^{2})\times\biggl(\omega^{2}-\Omega^{2}-\frac{1}{3}v_{Fe}^{2}\biggl(1-\frac{1}{10}\frac{v_{Fe}^{2}}{c^{2}}\biggr)k_{x}^{2}$$
\begin{equation}\label{E-P_QP Tr disp eq without spin Ey}+\frac{\pi e^{2}\hbar^{2}}{m^{3}c^{2}}n_{0}k_{x}^{2}-\frac{\hbar^{2}k_{x}^{4}}{4m^{2}}-\Xi^{2} k_{x}^{2}\biggr)-2\omega_{Le}^{2}\Omega^{2}=0.\end{equation}
Simultaneous account of the Darwin and annihilation interaction shifts the Fermi velocity. As we can see from terms in the big brackets in equation (\ref{E-P_QP Tr disp eq without spin Ey}). $\Xi^{2}$ is presented by formula (\ref{E-P_QP Xi def}).

Presence of magnetic moment of particles in equation for $\delta E_{y}$ changes the last term in dispersion equation (\ref{E-P_QP Tr disp eq without spin Ey})
$$(\omega^{2}-k_{x}^{2}c^{2}-2\omega_{Le}^{2})\times$$
$$\times\biggl(\omega^{2}-\Omega^{2}+\frac{\pi e^{2}\hbar^{2}}{m^{3}c^{2}}n_{0}k_{x}^{2}-\frac{1}{3}v_{Fe}^{2}\biggl(1-\frac{1}{10}\frac{v_{Fe}^{2}}{c^{2}}\biggr)k_{x}^{2}$$
\begin{equation}\label{E-P_QP Tr disp eq with spin Ey} -\frac{\hbar^{2}k_{x}^{4}}{4m^{2}}-\Xi^{2} k_{x}^{2}\biggr) -2\omega_{Le}^{2}\Omega^{2}\biggl(1+\frac{\mu_{0} k_{x}^{2}c}{e\Omega}\biggr)=0.\end{equation}

Equation for $E_{z}$ describes transverse waves with the following dispersion equation
$$\omega^{2}-k_{x}^{2}c^{2}-2\omega_{Le}^{2}$$
\begin{equation}\label{E-P_QP Tr DE Ez} -8\pi n_{0}\mu_{0}\frac{2\gamma}{\hbar}\frac{k_{x}^{2}c^{2}(\Theta+\Lambda)}{\omega^{2}-\Theta^{2}+\Lambda^{2}}=0. \end{equation}
Far from resonance frequency of the last term the spin gives small contribution in dispersion of the electromagnetic waves. Hence we can calculate contribution of spin in the electromagnetic wave dispersion by iteration method. Thus we obtain
$$\omega^{2}_{EM}=2\omega_{Le}^{2}+k_{x}^{2}c^{2}$$
\begin{equation}\label{E-P_QP Tr EM wave with spin Ez} +8\pi n_{0}\mu_{0}\frac{2\gamma}{\hbar}\frac{k_{x}^{2}c^{2}(\Theta+\Lambda)}{2\omega_{Le}^{2}+k_{x}^{2}c^{2}-\Theta^{2}+\Lambda^{2}}. \end{equation}
Analogous solution for electron-ion plasmas was found in Ref. \cite{Andreev VestnMSU 2007} (see formula 8). The last term in formula (\ref{E-P_QP Tr EM wave with spin Ez}) is the contribution of spin (equilibrium magnetization).

We can consider approximate expression for formula (\ref{E-P_QP Tr EM wave with spin Ez}) at small magnetization
$$\omega^{2}_{EM}=2\omega_{Le}^{2}+k_{x}^{2}c^{2}$$
\begin{equation}\label{E-P_QP Tr EM wave with spin Ez small magn} +8\pi n_{0}\mu_{0}B_{0}\biggl(\frac{2\gamma}{\hbar}\biggr)^{2}\frac{k_{x}^{2}c^{2}}{2\omega_{Le}^{2}+k_{x}^{2}c^{2}-(\frac{2\gamma}{\hbar})^{2}}. \end{equation}

In the limit of large annihilation interaction contribution, corresponding to the large magnetization we find the following spectrum of the electromagnetic wave, we also assume long wavelength limit,
$$\omega^{2}_{EM}=2\omega_{Le}^{2}+k_{x}^{2}c^{2}$$
\begin{equation}\label{E-P_QP Tr EM wave with spin Ez large annih} +2\biggl(\frac{2\gamma}{\hbar}\biggr)^{2}\frac{(4\pi)^{2} (n_{0}\mu_{0})^{2} k_{x}^{2}c^{2}}{2\omega_{Le}^{2}+k_{x}^{2}c^{2}-4\pi\frac{2\gamma}{\hbar}n_{0}\mu_{0}\Omega_{\gamma}}. \end{equation}

\subsection{Spin-plasma waves}

When contribution of the magnetization in equation (\ref{E-P_QP Tr DE Ez}) is small we can find $\omega(k)$ of a spin-plasma wave. This solution exists at frequencies close to the resonance frequency of the last term in formula (\ref{E-P_QP Tr DE Ez}) $\omega\approx\sqrt{\Theta^{2}-\Lambda^{2}}$. We consider solution in the following form $\omega=\sqrt{\Theta^{2}-\Lambda^{2}}+\delta\omega$ assuming that $\delta\omega\ll\sqrt{\Theta^{2}-\Lambda^{2}}$. Consequently we find
$$\omega=\sqrt{\Theta^{2}-\Lambda^{2}}$$
\begin{equation}\label{E-P_QP spin wave perpendicular}  -\frac{2\gamma}{\hbar}\frac{\Theta+\Lambda}{\sqrt{\Theta^{2}-\Lambda^{2}}}\frac{4\pi n_{0}\mu_{0}k_{x}^{2}c^{2}}{2\omega_{Le}^{2}+k_{x}^{2}c^{2}+\Lambda^{2}-\Theta^{2}}.\end{equation}
Similar solution for electron-ion plasmas was found in 2007 in Ref. \cite{Andreev VestnMSU 2007} (see formula (9)), see also Refs. \cite{Andreev AtPhys 08} (the formula before formula (16)) and \cite{Andreev IJMP 12}.

It was found that in electron-ion plasmas there are two spin-plasma waves located near the electron cyclotron frequency and ion cyclotron frequency \cite{Andreev VestnMSU 2007}. Since module of the cyclotron frequencies of electrons and positrons equal to each other we have found one spin-plasma wave for electron-positron plasmas.

Solutions obtained in Refs. \cite{Andreev VestnMSU 2007}, \cite{Andreev AtPhys 08}, \cite{Andreev IJMP 12} do not contain contributions the anomalous magnetic dipole moment, the quantum part of the spin current (quantum Bohm potential in the spin evolution equation). Previous papers \cite{Andreev VestnMSU 2007}, \cite{Andreev AtPhys 08}, \cite{Andreev IJMP 12} do not contain contribution of the Darwin and annihilation interactions as well.

The annihilation interaction changes the dispersion dependencies of waves propagating perpendicular to the external magnetic field for spin-1/2 quantum plasmas. The resonance frequency for spin-plasma waves propagating perpendicular to the external magnetic field is the same as for the spin-plasma waves propagating parallel to the external magnetic field. Formulas (\ref{E-P_QP omega Res small M})-(\ref{E-P_QP omega Res large M}) and Figs. (\ref{FD_SW1})-(\ref{FD_SW4}) are correct for the problem under consideration.

Let us consider formula (\ref{E-P_QP spin wave perpendicular}) in limits of small magnetization and long wavelengths
$$\omega=\frac{2\gamma}{\hbar}B_{0}+\frac{\hbar\mu_{0}}{2m\gamma}k_{x}^{2}+\frac{4\pi\gamma}{\hbar}n_{0}\mu_{0}$$
\begin{equation}\label{E-P_QP} -\frac{2\gamma}{\hbar}\biggl(1+\frac{2\gamma}{\hbar}\frac{4\pi n_{0}\mu_{0}}{B_{0}}\biggr)\frac{4\pi n_{0}\mu_{0}k_{x}^{2}c^{2}}{2\omega_{Le}^{2}+k_{x}^{2}c^{2}+(\frac{2\gamma}{\hbar}B_{0})^{2}}. \end{equation}
Shifts of frequency, from the cyclotron frequency $\frac{2\gamma}{\hbar}B_{0}$, caused by the quantum Bohm potential and annihilation interaction are positive. The first three terms form the resonance frequency. The shift of frequency from the resonance frequency is negative. Its module increases with increasing of the wave vector $k_{x}^{2}$.

Area of large magnetization and short wavelengths gives contribution via the quantum Bohm potential term and the annihilation term in the resonance frequency. Their behavior is depicted on Figs. (\ref{FD_SW1})-(\ref{FD_SW4}) and discussed in previous section.

\section{Oblique propagation of longitudinal waves}

\begin{figure}
\includegraphics[width=8cm,angle=0]{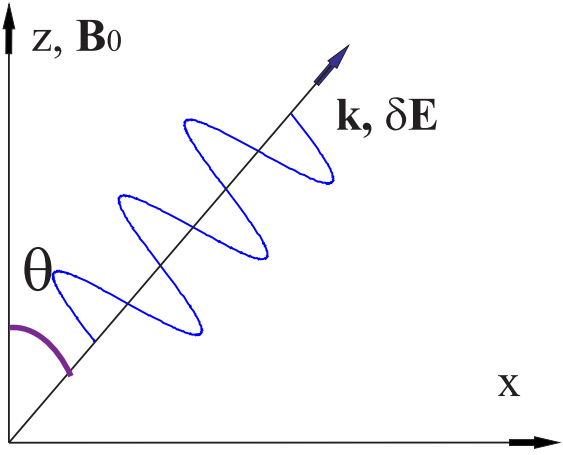}
\caption{\label{FD_oblPr} (Color online) The figure shows oblique propagation of a longitudinal wave in magnetized plasmas.}
\end{figure}

In this section we focus our attention on the longitudinal waves in electron-positron plasmas. We consider propagation of longitudinal waves at arbitrary angle to the external magnetic field (see Fig. (\ref{FD_oblPr})).

As solution of the QHD equations in the linear approximation we obtain the following dispersion equation
\begin{equation}\label{E-P_QP DE obliq full} 1-\frac{2\omega_{Le}^{2}\Re}{\omega^{2}(\omega^{2}-\Omega^{2})-\Re U^{2}k^{2}}=0,\end{equation}
where $k^{2}=k_{x}^{2}+k_{z}^{2}$, $\Xi$ is presented by formula (\ref{E-P_QP Xi def}),
\begin{equation}\label{E-P_QP} \Re\equiv\omega^{2}\sin^{2}\theta+(\omega^{2}-\Omega^{2})\cos^{2}\theta,\end{equation}
and
\begin{equation}\label{E-P_QP} U^{2}\equiv\frac{v_{Fe}^{2}}{3}\biggl(1-\frac{1}{10}\frac{v_{Fe}^{2}}{c^{2}}\biggr)-\frac{\pi e^{2}\hbar^{2}n_{0}}{m^{3}c^{2}}+\frac{\hbar^{2}k^{2}}{4m^{2}}-\Xi^{2}.\end{equation}

In limit cases of the parallel and the perpendicular propagation of waves we have one solution, which is the Langmuir wave. These solutions coincide with results obtained above (\ref{E-P_QP DE longit general Lang 2}) and (\ref{E-P_QP Langmuir wave perpendicular}).

At intermediate angles equation (\ref{E-P_QP DE obliq full}) gives two solutions. Let us consider contribution of the Darwin and annihilation interactions in spectrum of the oblique Langmuir wave and spectrum of the second branch of dispersion dependence.

Solving equation (\ref{E-P_QP DE obliq full}) we find explicit form of two branches of dispersion dependence
\begin{equation}\label{E-P_QP Langmuir obliq} \omega^{2}=\Upsilon+\sqrt{\Upsilon^{2}-(2\omega_{Le}^{2}+U^{2}k^{2})^{2}\cos^{2}\theta},\end{equation}
and
\begin{equation}\label{E-P_QP DE wave longit obliq} \omega^{2}=\frac{2\omega_{Le}^{2}+U^{2}k^{2}}{\Upsilon+\sqrt{\Upsilon^{2}-(2\omega_{Le}^{2}+U^{2}k^{2})^{2}\cos^{2}\theta}}\Omega^{2}\cos^{2}\theta, \end{equation}
where $\Upsilon=\omega_{Le}^{2}+\frac{1}{2}k^{2}c^{2}+\frac{1}{2}\Omega^{2}$. The expression under the square root can be rewritten as $(2\omega_{Le}^{2}+U^{2}k^{2}-\Omega^{2})^{2}+(2\omega_{Le}^{2}+U^{2}k^{2})\Omega^{2}\cos^{2}\theta$. The first of these solutions (\ref{E-P_QP Langmuir obliq}) is the Langmuir wave propagating at angle $\theta$ to direction of the external magnetic field. The second branch of the longitudinal waves (\ref{E-P_QP DE wave longit obliq}) disappears at $\theta=\pi/2$. At $\theta=0$ we have formal solution $\omega^{2}=\Omega^{2}$ from formula (\ref{E-P_QP DE wave longit obliq}). But we know from more general equation (\ref{E-P_QP DE obliq full}) that these is no such solution. Formula (\ref{E-P_QP DE wave longit obliq}) exists only intermediate angles.

From formulas (\ref{E-P_QP Langmuir obliq}) and (\ref{E-P_QP DE wave longit obliq}) and explicit form of $U^{2}$ and $\Xi$ we see that the Darwin and annihilation interactions lead to decreasing of $U^{2}$. As a consequence we find that these interactions decrease square of frequencies.

\section{Further generalization of quantum hydrodynamics of electron-positron plasmas}

Above we have developed the method of many-particle quantum
hydrodynamics for electron-positron plasmas. We have included a
necessary part of electron-positron interaction. It is the
annihilation interaction (see for instance theory of the
positronium \cite{Lewis PRA 73}). We have also presented some
applications of the annihilation interaction to magnetized plasma
wave dispersion.

Hydrodynamics is one of methods of plasma evolution description.
Kinetic equation play significant role at plasma description
\cite{Wigner PR 84}. Here we present quantum kinetics of
electron-positron plasmas. This kinetics is derived by means the
method, which directly follows from general conception of
many-particle hydrodynamics \cite{Andreev kinetics 12},
\cite{Andreev kinetics 13}.

Different approaches to quantum kinetics are presented in
literature \cite{Vagin Izv RAN 06}, \cite{Polyakov}, \cite{Kuzelev
Ruhadze UFN1999}, \cite{Kuzelev PhUsp 11}, \cite{Maksimov TMP
2002}. Recent application of Wigner kinetics have been also
presented in literature \cite{Zamanian PP 10}, \cite{Marklund PL A
11}, \cite{Zamanian NJP 10}.  Extensions of formalism to include
nonlocal quantum behavior via the Bohm potential were presented in
Refs. \cite{Schmidt PRE 12}, \cite{Tsintsadze EPL 09}.
Quasi-classic kinetic equations in extended nine dimensional phase
space, where the distribution function is considered as a function
of coordinate $\textbf{r}$, momentum $\textbf{p}$ and spin
$\textbf{s}$ \cite{Brodin PRL 08}, \cite{Brodin PRE 13}.

To include the quantum Bohm potential in kinetics authors of Ref.
\cite{Tsintsadze EPL 09} considered the quantum Bohm potential as
an extra interaction and put it as addition to the Lorentz force.
We should point out that the quantum Bohm potential is not an
interaction force. It appears from the flux of the momentum
density. Contribution of the quantum Bohm potential does not
appear in solutions of kinetic equations derived below.
Nevertheless, integrating the kinetic equations we get equations
of many-particle QHD presented above, which contain the quantum
Bohm potential. We conclude that the quantum Bohm potential is
hidden in the quantum kinetic equations presented below, but we do
not have a method to extract it. One needs to analyse quantum
kinetic theory to find the quantum Bohm potential contribution,
instead of forcing its contribution into the theory.

Physical variables in quantum mechanics appear as average of corresponding operator. Thus, if we need to derive kinetic equations we should present definition of the distribution function. Most famous quantum distribution function was suggested by Wigner \cite{Wigner PR 84}, but we do not use it. We start with classic microscopic distribution function \cite{Weinberg book}, \cite{Klimontovich book}. We change classic dynamic functions of position and momentum of particles on the corresponding operators. As the result we find the operator of many-particle microscopic quantum distribution function \cite{Andreev kinetics 12},
\cite{Andreev kinetics 13}
\begin{equation}\label{E-P_QP} \hat{f}=\sum_{i}\delta(\textbf{r}-\widehat{\textbf{r}}_{i})\delta(\textbf{p}-\widehat{\textbf{p}}_{i}).\end{equation}

Quantum mechanical averaging of the operator of many-particle distribution function gives us the microscopic distribution function for system of spinning particles \cite{Andreev kinetics 12},
\cite{Andreev kinetics 13}
$$f_{a}(\textbf{r}, \textbf{p},t)=\frac{1}{4}\int \Biggl(\psi^{*}(R,t)\sum_{i}\biggl(\delta(\textbf{r}-\textbf{r}_{i})\delta(\textbf{p}-\widehat{\textbf{p}}_{i})$$
\begin{equation}\label{E-P_QP def distribution function} +\delta(\textbf{p}-\widehat{\textbf{p}}_{i})\delta(\textbf{r}-\textbf{r}_{i})\biggr)\psi(R,t)+h.c.\Biggr)dR,\end{equation}
for each species of particles, i.e. electrons and positrons.

Differentiating distribution function of electrons with respect to time and using the Schrodinger equation (\ref{E-P_QP Schrodinger eq}) with the many-particle Hamiltonian (\ref{E-P_QP Hamiltonian}) we obtain the quantum kinetic equation for subsystem of electrons in electron-positron plasmas
$$\partial_{t}f_{e}+\frac{\textbf{p}}{m}\partial_{\textbf{r}}f_{e}+\biggl(q_{e}\textbf{E}+\frac{q_{e}}{mc}[\textbf{p},\textbf{B}]\biggr)\partial_{\textbf{p}}f_{e}$$
$$+\Biggl(q_{e}^{2}\frac{\pi\hbar^{2}}{m^{2}c^{2}}\nabla\int d\textbf{p}'f_{e}(\textbf{r},\textbf{p}',t)$$
$$+\frac{5}{2}q_{e}q_{p}\frac{\pi\hbar^{2}}{m^{2}c^{2}}\nabla\int d\textbf{p}'f_{p}(\textbf{r},\textbf{p}',t)\Biggr)\partial_{\textbf{p}}f_{e}$$
$$+\gamma_{e}\partial_{\alpha} B^{\beta}(\textbf{r},t) \cdot\partial_{\textbf{p}\alpha} S_{e}^{\beta}(\textbf{r}, \textbf{p},t)$$
\begin{equation}\label{E-P_QP kinetic equation fe}
+2\pi\gamma_{e}\gamma_{p}\partial_{\textbf{r}\alpha}\int
d\textbf{p}' S_{p}^{\beta}(\textbf{r},\textbf{p}',t) \cdot
\partial_{\textbf{p}\alpha} S_{e}^{\beta}(\textbf{r},
\textbf{p},t)=0,\end{equation} where we have applied the
self-consistent field approximation and neglected quantum
contribution in term describing interparticle interaction (see
equations 9-12 in Ref. \cite{Andreev kinetics 13}).

Let us remind that $\gamma_{a}$ is the gyromagnetic ratio for
particles of species $a$. For electrons and positrons we have
$\mid\gamma_{a}\mid=g\mu_{B}$, where $\mu_{B}=\frac{e\hbar}{2mc}$
is the Bohr magneton, and $g=1.00116$ includes contribution of the
anomalous magnetic dipole moment of electrons and positrons.

New function in phase space has arose in equation $\textbf{S}_{e}(\textbf{r}, \textbf{p},t)$ and $\textbf{S}_{p}(\textbf{r}, \textbf{p},t)$. They are vector spin distribution functions of electrons and positrons.

In the same way we can derive the quantum kinetic equation for positron distribution function
$$\partial_{t}f_{p}+\frac{\textbf{p}}{m}\partial_{\textbf{r}}f_{p}+\biggl(q_{p}\textbf{E}+\frac{q_{p}}{mc}[\textbf{p},\textbf{B}]\biggr)\partial_{\textbf{p}}f_{p}$$
$$+\Biggl(q_{p}^{2}\frac{\pi\hbar^{2}}{m^{2}c^{2}}\nabla\int d\textbf{p}'f_{p}(\textbf{r},\textbf{p}',t)$$
$$+\frac{5}{2}q_{e}q_{p}\frac{\pi\hbar^{2}}{m^{2}c^{2}}\nabla\int d\textbf{p}'f_{e}(\textbf{r},\textbf{p}',t)\Biggr)\partial_{\textbf{p}}f_{p}$$
$$+\gamma_{p}\partial_{\alpha} B^{\beta}(\textbf{r},t) \cdot\partial_{\textbf{p}\alpha} S_{p}^{\beta}(\textbf{r}, \textbf{p},t)$$
\begin{equation}\label{E-P_QP kinetic equation fp}
+2\pi\gamma_{e}\gamma_{p}\partial_{\textbf{r}\alpha}\int
d\textbf{p}' S_{e}^{\beta}(\textbf{r},\textbf{p}',t) \cdot
\partial_{\textbf{p}\alpha} S_{p}^{\beta}(\textbf{r},
\textbf{p},t)=0. \end{equation}

Different terms in equations (\ref{E-P_QP kinetic equation fe}) and (\ref{E-P_QP kinetic equation fp}) have the following meaning.  The first and second terms are substantial derivatives of distribution functions. The third terms describe action of the Lorentz force on evolution of distribution functions. Other terms also present forces of interaction. The fourth term in equation (\ref{E-P_QP kinetic equation fe}) (equation (\ref{E-P_QP kinetic equation fp})) describes the Darwin interaction between electrons (between positrons). The fifth terms in equations (\ref{E-P_QP kinetic equation fe}) and (\ref{E-P_QP kinetic equation fp}) present superposition of the Darwin and the spinless part of annihilation interactions between electrons and positrons. Coefficient 5/2 consists of two parts, 1 appears from the Darwin interaction and 3/2 corresponds to the annihilation interaction. The sixth terms correspond to the spin-spin interaction and interaction of spins with the external magnetic field. The sixth terms include the spin-distribution function $\textbf{S}_{a}(\textbf{r}, \textbf{p},t)$ of species $a$.  The last terms describe the spin dependent part of the annihilation interaction between electrons and positrons.

The spin distribution functions for each species appears as the quantum mechanical average of the corresponding operator
\begin{equation}\label{E-P_QP} \hat{S}^{\alpha}=\sum_{i}\delta(\textbf{r}-\widehat{\textbf{r}}_{i})\delta(\textbf{p}-\widehat{\textbf{p}}_{i})\widehat{\sigma}^{\alpha}_{i}.\end{equation}
Hence explicit form of the spin distribution function is
$$S_{a}^{\alpha}(\textbf{r}, \textbf{p},t)=\frac{1}{4}\int \Biggl(\psi^{*}(R,t)\sum_{i}\biggl(\delta(\textbf{r}-\textbf{r}_{i})\delta(\textbf{p}-\widehat{\textbf{p}}_{i})$$
\begin{equation}\label{E-P_QP def spin distribution function}+\delta(\textbf{p}-\widehat{\textbf{p}}_{i})\delta(\textbf{r}-\textbf{r}_{i})\biggr)\widehat{\sigma}^{\alpha}_{i}\psi(R,t)+h.c.\Biggr)dR.\end{equation}
We present explanation of structure of the spin distribution function in physical terms. Nevertheless the definition appears, without any efforts, during derivation of kinetic equations (\ref{E-P_QP kinetic equation fe}) and (\ref{E-P_QP kinetic equation fp}).

Let us represent the kinetic equation for distribution function $f(\textbf{r}, \textbf{p}, t)$ in a brief form, which allows to get structure of the equation
\begin{equation}\label{E-P_QP} \partial_{t}f_{a}+\frac{\textbf{p}}{m}\partial_{\textbf{r}}f_{a}+\textbf{F}_{a} \partial_{\textbf{a}}f_{p}+\gamma_{a}\partial_{\alpha} B^{\beta}_{a,eff} \cdot\partial_{\textbf{p}\alpha} S_{a}^{\beta}=0,\end{equation}
where $\textbf{F}_{a}(\textbf{r},t)$ is the force field acting on the species $a$, and $\textbf{B}_{a,eff}(\textbf{r},t)$ is the effective magnetic field acting on magnetic moments of particles of species $a$.

Differentiating the spin distribution function of electrons with respect to time, using Schrodinger equation (\ref{E-P_QP Schrodinger eq}) with the Hamiltonian (\ref{E-P_QP Hamiltonian}) and applying same assumption as at derivation of equations (\ref{E-P_QP kinetic equation fe}) and (\ref{E-P_QP kinetic equation fp}) we find
$$\partial_{t}S_{e}^{\alpha}+\frac{\textbf{p}}{m}\partial_{\textbf{r}}S_{e}^{\alpha}+\biggl(q_{e}\textbf{E}+\frac{q_{e}}{mc}[\textbf{p},\textbf{B}]\biggr)\partial_{\textbf{p}}S_{e}^{\alpha}$$
$$+\Biggl(q_{e}^{2}\frac{\pi\hbar^{2}}{m^{2}c^{2}}\nabla\int d\textbf{p}'f_{e}(\textbf{r},\textbf{p}',t)$$
$$+\frac{5}{2}q_{e}q_{p}\frac{\pi\hbar^{2}}{m^{2}c^{2}}\nabla\int d\textbf{p}'f_{p}(\textbf{r},\textbf{p}',t)\Biggr)\partial_{\textbf{p}}S_{e}^{\alpha}$$
$$+\gamma_{e}\partial_{\beta} B^{\alpha}(\textbf{r},t)\partial_{\textbf{p}\beta} f_{e}-\frac{2\gamma_{e}}{\hbar}\varepsilon^{\alpha\beta\gamma}S_{e}^{\beta}B^{\gamma}$$
$$+2\pi\gamma_{e}\gamma_{p}\partial_{\beta} \int d\textbf{p}' S_{p}^{\alpha}(\textbf{r},\textbf{p}',t)\cdot\partial_{\textbf{p}\beta} f_{e}$$
\begin{equation}\label{E-P_QP kinetic equation Se} -2\pi\gamma_{p}\frac{2\gamma_{e}}{\hbar}\varepsilon^{\alpha\beta\gamma}S_{e}^{\beta}\int d\textbf{p}' S_{p}^{\gamma}(\textbf{r},\textbf{p}',t)=0.\end{equation}

We need to derive one more kinetic equation. It is equation for the spin distribution function of positrons. Applying the method described above we obtain
$$\partial_{t}S_{p}^{\alpha}+\frac{\textbf{p}}{m}\partial_{\textbf{r}}S_{p}^{\alpha}+\biggl(q_{p}\textbf{E}+\frac{q_{p}}{mc}[\textbf{p},\textbf{B}]\biggr)\partial_{\textbf{p}}S_{p}^{\alpha}$$
$$+\Biggl(q_{p}^{2}\frac{\pi\hbar^{2}}{m^{2}c^{2}}\nabla\int d\textbf{p}'f_{p}(\textbf{r},\textbf{p}',t)$$
$$+\frac{5}{2}q_{e}q_{p}\frac{\pi\hbar^{2}}{m^{2}c^{2}}\nabla\int d\textbf{p}'f_{e}(\textbf{r},\textbf{p}',t)\Biggr)\partial_{\textbf{p}}S_{p}^{\alpha}$$
$$+\gamma_{p}\partial_{\beta} B^{\alpha}(\textbf{r},t)\partial_{\textbf{p}\beta} f_{p} -\frac{2\gamma_{p}}{\hbar}\varepsilon^{\alpha\beta\gamma}S_{p}^{\beta}B^{\gamma}$$
$$+2\pi\gamma_{e}\gamma_{p}\partial_{\beta} \int d\textbf{p}' S_{e}^{\alpha}(\textbf{r},\textbf{p}',t)\cdot\partial_{\textbf{p}\beta} f_{p}$$
\begin{equation}\label{E-P_QP kinetic equation Sp} -2\pi\gamma_{e}\frac{2\gamma_{p}}{\hbar}\varepsilon^{\alpha\beta\gamma}S_{p}^{\beta}\int d\textbf{p}' S_{e}^{\gamma}(\textbf{r},\textbf{p}',t)=0.\end{equation}
We describe now physical meaning of different terms in the quantum kinetic equations for spin distribution functions.   Physical meaning of different terms in these equations are similar to physical meaning of terms in equations (\ref{E-P_QP kinetic equation fe}) and (\ref{E-P_QP kinetic equation fp}). The first and second terms in equations (\ref{E-P_QP kinetic equation Se}) and (\ref{E-P_QP kinetic equation Sp}) gives the substantial derivatives $(\partial_{t}+\textbf{v}\nabla)$ of the spin-distribution functions. The third terms are the Lorentz force.  The fourth terms are the electron-electron and positron-positron Darwin interactions. The fifth terms describe the interspecies Darwin and spinless part of annihilation interactions. The sixth and seventh terms in both equations describe spin-spin interaction. The last two terms in equations (\ref{E-P_QP kinetic equation Se}) and (\ref{E-P_QP kinetic equation Sp}) present spin dependent part of the annihilation interaction.

Set of kinetic equations is considered in the self-consistent field approximation and coupled with the Maxwell equations (\ref{E-P_QP div E})-(\ref{E-P_QP rot B}).

The kinetic equation for spin-distribution function has the following structure
$$\partial_{t}\textbf{S}_{a}+\frac{\textbf{p}}{m}\partial_{\textbf{r}}\textbf{S}_{a}+(\textbf{F}_{a}\partial_{\textbf{p}})\textbf{S}_{a}+\gamma_{a}\partial_{\beta} \textbf{B}_{a,eff}\partial_{\textbf{p}\beta}f_{a}$$
\begin{equation}\label{E-P_QP} -\frac{2\gamma_{a}}{\hbar}[\textbf{S}_{a}, \textbf{B}_{a,eff}]=0.\end{equation}

Kinetic equations look rather huge. This is due to the fact that we presented new terms in explicit form, which is an integral form. Applying hydrodynamic variables these equations can be represented.

Let us mention that the distribution functions presented in this section correspond to the hydrodynamic functions applied above. Hydrodynamic functions arise as moments of distribution functions: $n_{a}(\textbf{r},t)=\int d\textbf{p}f_{a}(\textbf{r},\textbf{p},t)$, $\textbf{j}_{a}(\textbf{r},t)=\frac{1}{m}\int d\textbf{p}\textbf{p}f_{a}(\textbf{r},\textbf{p},t)$

Let us consider the spin distribution function
\begin{equation}\label{E-P_QP def spin density} S_{a}^{\alpha}(\textbf{r},t)=\int dR\sum_{i}\delta(\textbf{r}-\textbf{r}_{i})\psi^{*}(R,t)\widehat{\sigma}^{\alpha}_{i}\psi(R,t),\end{equation}
proportional to the magnetization $\textbf{M}_{a}(\textbf{r},t)$ (\ref{E-P_QP def of magnetization}), usually used in the
quantum hydrodynamics \cite{MaksimovTMP 2001}, \cite{Andreev spin current}, and \cite{Andreev IJMP 12}: $\textbf{M}_{a}(\textbf{r},t)=\gamma_{a}\textbf{S}_{a}(\textbf{r},t)$. Next integrating the spin distribution function over the momentum we get the spin density appearing in the quantum hydrodynamic equations \cite{Andreev Asenjo 13}, \cite{Andreev spin current}, \cite{Andreev IJMP 12}
\begin{equation}\label{E-P_QP def spin density connection of S and S} S_{a}^{\alpha}(\textbf{r}, t)=\int S_{a}^{\alpha}(\textbf{r}, \textbf{p},t)d\textbf{p}.\end{equation}

\subsubsection{Application of quantum kinetics to dispersion of waves in electron-positron quantum plasmas}

We demonstrate application of quantum kinetics of electron-positron plasmas, which includes the Darwin and the annihilation interactions, to dispersion properties of waves in plasmas with no external field.

$$-\imath(\omega-\textbf{k}\textbf{v})\delta f_{e}-e(\delta \textbf{E}+[\textbf{v},\delta \textbf{B}]/c)\partial_{\textbf{p}}f_{0e}$$
\begin{equation}\label{E-P_QP} +\frac{\pi e^{2}\hbar^{2}}{m^{2}c^{2}}\biggl(\nabla\int \delta f_{e}d\textbf{p}'-\frac{5}{2}\nabla\int \delta f_{p}d\textbf{p}'\biggr) \partial_{\textbf{p}}f_{0e},\end{equation}
and
$$-\imath(\omega-\textbf{k}\textbf{v})\delta f_{p}+e(\delta \textbf{E}+[\textbf{v},\delta \textbf{B}]/c)\partial_{\textbf{p}}f_{0p}$$
\begin{equation}\label{E-P_QP} +\frac{\pi e^{2}\hbar^{2}}{m^{2}c^{2}}\biggl(\nabla\int \delta f_{p}d\textbf{p}'-\frac{5}{2}\nabla\int \delta f_{e}d\textbf{p}'\biggr) \partial_{\textbf{p}}f_{0p},\end{equation}
where we have used $q_{e}=-e$ and $q_{p}=e$.

Below we use $f_{0e}(\textbf{p})=f_{0p}(\textbf{p})\equiv
f_{0}(\textbf{p})$. After some calculations we find the
longitudinal dielectric constant $\varepsilon_{l}$, which appears
as
\begin{equation}\label{E-P_QP diel funct longitud} \varepsilon_{l}=1+\frac{8\pi e^{2}}{\omega k^{2}} \biggl(1+\frac{7}{2}\frac{\pi e^{2}\hbar^{2}}{m^{2}c^{2}}\aleph\biggr) \int\frac{(\textbf{k}\textbf{v})^{2}}{\omega-\textbf{k}\textbf{v}}\frac{\partial f_{0}}{\partial \epsilon}d\textbf{v}, \end{equation}
where $\epsilon=\frac{\textbf{p}^{2}}{2m}$ is the energy, and
\begin{equation}\label{E-P_QP}\aleph=\frac{\Im\biggl(1+\frac{3}{2}\frac{\pi e^{2}\hbar^{2}}{m^{2}c^{2}}\Im\biggr)}{1-2 \frac{\pi e^{2}\hbar^{2}}{m^{2}c^{2}}\Im-\frac{21}{4} \biggl(\frac{\pi e^{2}\hbar^{2}}{m^{2}c^{2}}\biggr)^{2}\Im^{2}}\approx\Im,\end{equation}
with
\begin{equation}\label{E-P_QP} \Im\equiv\int\frac{\textbf{k}\textbf{v}}{\omega-\textbf{k}\textbf{v}}\frac{\partial f_{0}}{\partial \epsilon} d\textbf{v}. \end{equation}

We obtain that the transverse dielectric constant $\varepsilon_{tr}$ does not change in compare with the case when the annihilation interaction is absent.

Dispersion equation has form of $\varepsilon_{l}=0$.

%%%%%%%%%%%%%%%%%

To get an explicit form of dispersion equation we should use an explicit form of the equilibrium distribution function $f_{0}(\textbf{p})$. We consider the Maxwell distribution function for equilibrium distribution
\begin{equation}\label{E-P_QP maxwell distrib}f_{0}(\textbf{p})=\frac{n_{0}m^{3/2}}{(\sqrt{2\pi T})^{3}}\exp\biggl(-\frac{m\textbf{v}^{2}}{2T}\biggr),\end{equation}
where $T$ is the temperature in units of energy, and
$\textbf{p}=m\textbf{v}$. The Darwin and annihilation interactions
have semi-relativistic nature, but they  do not contain an
explicit contribution of momentum. Consequently we can apply
non-relativistic equilibrium distribution function. Obviously the
Maxwell distribution function satisfy the kinetic equations with
the Darwin and annihilation interactions.

For the Maxwell distribution in equilibrium state we meet the following integral in the dielectric function (\ref{E-P_QP diel funct longitud})
$$Z(\alpha)=\frac{1}{\sqrt{\pi}}\int_{-\infty}^{+\infty}\frac{\exp(-\xi^{2})}{\xi-\alpha}d\xi$$
\begin{equation}\label{E-P_QP Z func definition}=\frac{1}{\sqrt{\pi}}\biggl[ P\int_{-\infty}^{+\infty}\frac{\exp(-\xi^{2})}{\xi-\alpha}d\xi\biggr]+\imath\sqrt{\pi}\exp(-\alpha^{2}),\end{equation}
where $\alpha=\frac{\omega}{kv_{T}}$ with $v_{T}\equiv\sqrt{\frac{T}{m}}$, and the symbol $P$ denotes the principle part of the integral.

Let us present assumptions of formula (\ref{E-P_QP Z func definition}).
At $\alpha\gg 1$ we have
\begin{equation}\label{E-P_QP Z big alpha} Z(\alpha)\simeq-\frac{1}{\alpha}\biggl(1+\frac{1}{2\alpha^{2}}+\frac{3}{4\alpha^{4}}+...\biggr)+\imath\sqrt{\pi}\exp(-\alpha^{2}),\end{equation}

Finally dispersion equation (\ref{E-P_QP diel funct longitud}) reappear as
$$1+4\frac{\omega_{Le}^{2}}{k^{2}v_{T}^{2}}(1+\alpha Z(\alpha))\times$$
\begin{equation}\label{E-P_QP} \times\biggl(1+\frac{7}{2}\frac{\pi e^{2}\hbar^{2}}{m^{2}c^{2}}\frac{n_{0}}{T}(1+\alpha Z(\alpha))\biggr)=0,\end{equation}
or in more explicit form
$$1-2\frac{1}{\alpha^{2}}\frac{\omega_{Le}^{2}}{k^{2}v_{T}^{2}}-2\frac{1}{\alpha^{4}}\frac{\omega_{Le}^{2}}{k^{2}v_{T}^{2}} \biggl(3-\frac{7}{2}\frac{\pi e^{2}\hbar^{2}}{m^{3}c^{2}}\frac{n_{0}}{v_{T}^{2}}\biggr)$$
\begin{equation}\label{E-P_QP} +\imath\alpha\sqrt{\pi}2\frac{\omega_{Le}^{2}}{k^{2}v_{T}^{2}}\exp(-\alpha^{2}) \biggl(1-4\frac{7}{2}\frac{\pi e^{2}\hbar^{2}}{m^{3}c^{2}}\frac{n_{0}}{v_{T}^{2}}\frac{1}{\alpha^{2}}\biggr)=0. \end{equation}

\begin{figure}
\includegraphics[width=8cm,angle=0]{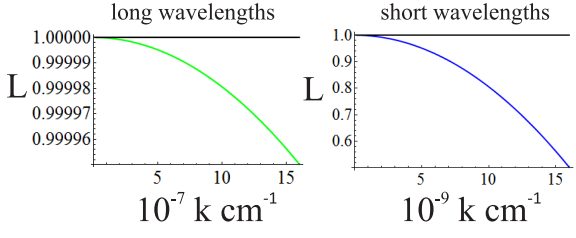}
\caption{\label{FD_FermiVsDarwin} (Color online) The figure shows
decreasing of the Landau damping of the Langmuir wave in quantum electron-positron plasmas due to
simultaneous contribution of the Darwin and annihilation
interactions. Contribution of interactions is presented via multiplier $L\equiv (1-\frac{7}{4} \frac{\hbar^{2}}{m^{2}c^{2}}k^{2})$ presented in the end of formula (\ref{E-P_QP Landau Damping}). Horizontal line $L=1$ shows multiplier $L$ in absence of the Darwin and annihilation
interactions.}
\end{figure}

Real part of frequency appears as
\begin{equation}\label{E-P_QP} \omega^{2}=2\omega_{Le}^{2}+\biggl(3 v_{T}^{2}-\frac{7}{2}\frac{\pi e^{2}\hbar^{2}}{m^{3}c^{2}}n_{0}\biggr)k^{2}. \end{equation}
For imaginary part we find
$$ Im\delta\omega=-\sqrt{\pi}\biggl(\frac{\sqrt{2}\omega_{Le}}{kv_{T}}\biggr)^{3}\exp\Biggl(-\frac{2\omega_{Le}^{2}}{k^{2}v_{T}^{2}}\Biggr)\times$$
\begin{equation}\label{E-P_QP Landau Damping} \times \frac{2\omega_{Le}^{2}}{k^{2}v_{T}^{2}}\Biggl(1-\frac{7}{4}\frac{\hbar^{2}}{m^{2}c^{2}}k^{2}\Biggr). \end{equation}
The full formula for frequency is $\omega=Re\omega+\imath Im\omega$.

Only spinless part of the annihilation interaction gives contribution in the Langmuir wave dispersion in absence of the external magnetic field. It comes together with the Darwin interaction. The terms containing all these interactions has coefficient $7/2$. This coefficient has the following structure. 1 appears from the Darwin interaction between electrons and between positrons. Another 1 comes from interspecies (electron-positron) interaction. 3/2 comes from the annihilation interaction between electrons and positrons.

We see that the Darwin and annihilation interactions decrease
frequency and damping rate of the Langmuir waves.

\section{NLSEs for electron-positron plasmas}

In approximation of the eddy-free motion we can associate set of QHD equations with a pair of non-linear spinor Schrodinger (Pauli) equations (NLSEs). We have a pair of equations since we consider plasmas consisting of two species: electron and positrons.

In case of spinless fermions or bosons we can directly derive a non-linear Schrodinger equation for the wave function in medium from the set of QHD equations (the continuity and Euler equations) \cite{MaksimovTMP 1999}, \cite{Andreev PRA08}. There has not been suggested a direct derivation of the NLSE for spinning particles. However, as we have mentioned, we can associate a NLSE with set of QHD equations for spinning particles (the continuity, Euler, and magnetic moment evolution equations).

For quantum electron-positron plasmas with the Darwin and annihilation interactions the pair of NLSEs appears as
$$\imath\hbar\partial_{t}\Phi_{e}=\biggl(\frac{\textbf{D}_{e}^{2}}{2m}-e\varphi+(3\pi^{2})^{\frac{2}{3}}\frac{\hbar^{2}}{2m}n_{e}^{\frac{2}{3}}
-\gamma_{e}\textbf{B}\mbox{\boldmath $\sigma$}$$
\begin{equation}\label{E-P_QP NLSE electrons} -\frac{\pi e^{2}\hbar^{2}}{m^{2}c^{2}}n_{e}
+\frac{5}{2}\frac{\pi e^{2}\hbar^{2}}{m^{2}c^{2}}n_{p}-2\pi\textbf{M}_{p}\mbox{\boldmath $\sigma$}\biggr)\Phi_{e}, \end{equation}
and
$$\imath\hbar\partial_{t}\Phi_{p}=\biggl(\frac{\textbf{D}_{p}^{2}}{2m}+e\varphi+(3\pi^{2})^{\frac{2}{3}}\frac{\hbar^{2}}{2m}n_{p}^{\frac{2}{3}}
-\gamma_{p}\textbf{B}\mbox{\boldmath $\sigma$}$$
\begin{equation}\label{E-P_QP NLSE positrons} -\frac{\pi e^{2}\hbar^{2}}{m^{2}c^{2}}n_{p}
+\frac{5}{2}\frac{\pi e^{2}\hbar^{2}}{m^{2}c^{2}}n_{e}-2\pi\textbf{M}_{p}\mbox{\boldmath $\sigma$}\biggr)\Phi_{p}, \end{equation}
where $\textbf{D}_{a}=-\imath\hbar\nabla-\frac{q_{a}}{c}\textbf{A}(\textbf{r},t)$, $n_{a}=\Phi_{a}^{*}\Phi_{a}$, $\textbf{M}_{a}=\Phi_{a}^{*}\gamma_{a}\mbox{\boldmath $\sigma$}\Phi_{a}$.
Physical meaning of different terms in the NLSEs is analogous to physical meaning of terms in the Euler equations (\ref{E-P_QP Euler equation for electrons}) and (\ref{E-P_QP Euler equation for positrons}). The sixth terms in the right-hand side of equations (\ref{E-P_QP NLSE electrons}) and (\ref{E-P_QP NLSE positrons}) are combination of the interspecies Darwin interaction and spinless part of the annihilation interaction. In the Euler equation these terms are presented separately.

Set of NLSEs is obtained in the self-consistent field approximation and coupled with the Maxwell equations (\ref{E-P_QP div E})-(\ref{E-P_QP rot B}).

In spinless case the wave function in medium (the order parameter) for species $a$ is defined as follows
\begin{equation}\label{E-P_QP} \Phi_{a}^{SL}(\textbf{r},t)=\sqrt{n_{a}(\textbf{r},t)}\exp(\frac{\imath}{\hbar}m_{a}\phi_{a}(\textbf{r},t)),\end{equation}
where $\phi_{a}(\textbf{r},t)$ is the potential of the velocity field of species $a$.

For spinning particles the wave function in medium is the spinor function
\begin{equation}\label{E-P_QP} \Phi_{a}(\textbf{r},t)=\sqrt{n_{a}(\textbf{r},t)}\exp(\frac{\imath}{\hbar}m_{a}\phi_{a}(\textbf{r},t))\zeta(\textbf{r},t),\end{equation}
where $\zeta(\textbf{r},t)$ is the unit spinor function.

NLSE is an effective tool in various field of physics. Hence it is important to have it for electron-positron plasmas. It can be applied instead of the set of QHD equations.

\section{Vorticity of electron-positron quantum plasmas}

In this section we briefly describe modification of equations for the grand generalized
vorticities and the Clebsch potential vector fields associated with the grand generalized
vorticities due to presence of the Darwin and annihilation interactions. At description of vorticities we follow Ref. \cite{Mahajan PRL 11} describing contribution of new interactions.

\begin{figure}
\includegraphics[width=8cm,angle=0]{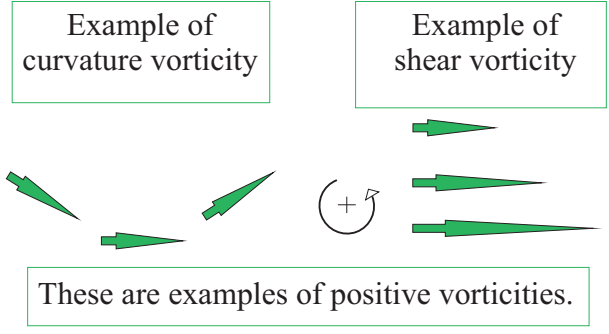}
\caption{\label{FD_Vort} (Color online) The figure illustrates
non-zero voticity in continuous mediums.}
\end{figure}

When we have deal with the degenerate electron-ion plasmas we can consider motionless ions giving positively charged background. In this case we study behavior of electrons and our QHD equations look like equations for a single fluid. Consequently a single vorticity can be introduced. For electron-positron plasmas it is no longer true. Now we need to introduce a pair of vorticities.

From Euler equation for species $a$ the classical generalized vorticity can be found
\begin{equation}\label{E-P_QP} \mbox{\boldmath $\Omega$}_{(a),c}=\textbf{B}+\frac{mc}{q_{a}}\nabla\times \textbf{v}_{a}. \end{equation}
The quantum vorticity for species $a$ appears from the corresponding magnetic moment evolution equation \cite{Mahajan PRL 11}, \cite{Takabayasi PTP 83}
\begin{equation}\label{E-P_QP} \mbox{\boldmath $\Omega$}_{(a),q}=\frac{\nabla \mu_{(a),x}\times\nabla\mu_{(a),y}}{\gamma_{a}\mu_{(a),z}}.\end{equation}
The quantum vorticity $\mbox{\boldmath $\Omega$}_{(a),q}$ can be rewritten in tensor  notations $\Omega^{\alpha}_{(a),q}=\frac{1}{2}\varepsilon^{\alpha\beta\gamma}\varepsilon^{\delta\mu\nu}\mu^{\delta}_{a}\partial^{\beta}\mu^{\mu}_{a}\partial^{\gamma}\mu^{\nu}_{a}$. Vorticity of a vector field is demonstrated on Fig. (\ref{FD_Vort}).

Equations for vorticity time evolution are derived from the hydrodynamic equations
\begin{equation}\label{E-P_QP vort evol Cl} \partial_{t}\mbox{\boldmath $\Omega$}_{(a),c}=\nabla\times(\textbf{v}_{a}\times\mbox{\boldmath $\Omega$}_{(a),c})+\frac{\hbar}{2m\gamma_{a}}\nabla\mu^{\beta}_{a}\times\nabla \hat{B}^{\beta}_{a}, \end{equation}
and
\begin{equation}\label{E-P_QP vort evol Q} \partial_{t}\mbox{\boldmath $\Omega$}_{(a),q}=\nabla\times(\textbf{v}_{a}\times\mbox{\boldmath $\Omega$}_{(a),q})+\frac{\hbar}{2m\gamma_{a}}\nabla\mu^{\beta}_{a}\times\nabla \hat{B}^{\beta}_{a}, \end{equation}
with the effective magnetic field
\begin{equation}\label{E-P_QP magn field eff e} \widehat{\textbf{B}}_{e}=\textbf{B}+\frac{\hbar c}{2q_{e}\gamma_{e}}\frac{1}{n_{e}}\partial^{\beta}(n_{e}\partial^{\beta}\mbox{\boldmath $\mu$}_{e})+2\pi n_{p}\mbox{\boldmath $\mu$}_{p},\end{equation}
and
\begin{equation}\label{E-P_QP magn field eff p} \widehat{\textbf{B}}_{p}=\textbf{B}+\frac{\hbar c}{2q_{p}\gamma_{p}}\frac{1}{n_{p}}\partial^{\beta}(n_{p}\partial^{\beta}\mbox{\boldmath $\mu$}_{p})+2\pi n_{e}\mbox{\boldmath $\mu$}_{e}.\end{equation}

Equations (\ref{E-P_QP vort evol Cl}) and (\ref{E-P_QP vort evol Q}) look almost the same. They describe time evolution of classic and quantum vorticities. Identical structure of these equations allows us to introduce the grand generalized
vorticities as sum or difference of the classic and the quantum vorticities.

Effective magnetic fields (\ref{E-P_QP magn field eff e}) and (\ref{E-P_QP magn field eff p}) appear in the Euler and magnetic moment evolution equations in terms describing the spin-spin interactions by including of non-potential parts of the quantum Bohm potential and spin-dependent part of the annihilation interaction. Effective magnetic field (\ref{E-P_QP magn field eff e}) ((\ref{E-P_QP magn field eff p})) arises in QHD equations describing evolution of electrons (positrons). Nevertheless it contains dependence on the concentration and the reduced magnetization of another species.

Spin-dependent part of the annihilation interaction brings extra terms in the Euler equations (\ref{E-P_QP Euler equation for electrons}), (\ref{E-P_QP Euler equation for positrons}) and the magnetic moment evolution equations (\ref{E-P_QP eq of magnetic moments evol electr}), (\ref{E-P_QP eq of magnetic moments evol positr}). These terms lead to contribution of positron magnetization in the quantum vorticity of electrons and vice-versa.

Contribution of the annihilation interaction changes the effective magnetic field, but it does not break structure of the vorticity equations.

The classical generalized vorticity and the quantum vorticity allow to get two grand generalized
vorticities for each species
\begin{equation}\label{E-P_QP} \mbox{\boldmath $\Omega$}_{(a),\pm}=\mbox{\boldmath $\Omega$}_{(a),c} \pm\frac{\hbar c}{2q_{a}\gamma_{a}}\mbox{\boldmath $\Omega$}_{(a),q}.\end{equation}

Applying equations (\ref{E-P_QP vort evol Cl}) and (\ref{E-P_QP vort evol Q}) we can easily obtain equations for the grand generalized
vorticities
\begin{equation}\label{E-P_QP vort + evol} \partial_{t}\mbox{\boldmath $\Omega$}_{(a),+}=\nabla\times(\textbf{v}_{a}\times\mbox{\boldmath $\Omega$}_{(a),+})+\frac{\hbar}{m\gamma_{a}}\nabla\mu^{\beta}_{a}\times\nabla \hat{B}^{\beta}_{a}, \end{equation}
and
\begin{equation}\label{E-P_QP vort - evol} \partial_{t}\mbox{\boldmath $\Omega$}_{(a),-}=\nabla\times(\textbf{v}_{a}\times\mbox{\boldmath $\Omega$}_{(a),-}), \end{equation}
where the last term in equation (\ref{E-P_QP vort + evol}) is the
source of vorticities. Equation (\ref{E-P_QP vort - evol}) shows
conservation of vorticities $\Omega_{(a),-}$ for each species
$a=e, p$, since it does not contain any source.

The Clebsch potential vector fields associated with the grand generalized
vorticities $\textbf{P}_{(a),\pm}=\textbf{P}_{(a),c} \pm \frac{\hbar c}{2q_{a}\gamma_{a}}\textbf{P}_{(a),q}$ satisfy
\begin{equation}\label{E-P_QP} \partial_{t}\textbf{P}_{(a),+}=\textbf{v}_{a}\times\mbox{\boldmath $\Omega$}_{(a),+} +\frac{\hbar}{m\gamma_{a}}\mu^{\beta}\nabla \hat{B}^{\beta}_{a}+\frac{c}{q_{a}}\mbox{\boldmath $\chi$}_{a}, \end{equation}
and
\begin{equation}\label{E-P_QP} \partial_{t}\textbf{P}_{(a),-}=\textbf{v}_{a}\times\mbox{\boldmath $\Omega$}_{(a),-} +\frac{c}{q_{a}}\mbox{\boldmath $\chi$}_{a},\end{equation}
where

$$\mbox{\boldmath $\chi$}_{e}=-\frac{\nabla p_{e}}{n_{e}}+\frac{\hbar^{2}}{2m}\nabla\biggl(\frac{\nabla^{2}\sqrt{n_{e}}}{\sqrt{n_{e}}}\biggr)$$
$$+\frac{\pi e^{2}\hbar^{2}}{m^{2}c^{2}}\nabla n_{e}-\frac{5}{2}\frac{\pi e^{2}\hbar^{2}}{m^{2}c^{2}}\nabla n_{p}$$
\begin{equation}\label{E-P_QP}
+\frac{\hbar^{2}}{8m\gamma_{e}^{2}}\nabla(\partial^{\beta}\mu^{\gamma}_{e}\cdot\partial^{\beta}\mu^{\gamma}_{e})
-\nabla(q_{e}\varphi+\frac{1}{2}m\textbf{v}_{e}^{2}) ,\end{equation}
and
$$\mbox{\boldmath $\chi$}_{p}=-\frac{\nabla p_{p}}{n_{p}}+\frac{\hbar^{2}}{2m}\nabla\biggl(\frac{\nabla^{2}\sqrt{n_{p}}}{\sqrt{n_{p}}}\biggr)$$
$$+\frac{\pi e^{2}\hbar^{2}}{m^{2}c^{2}}\nabla n_{p}-\frac{5}{2}\frac{\pi e^{2}\hbar^{2}}{m^{2}c^{2}}\nabla n_{e}$$
\begin{equation}\label{E-P_QP}
+\frac{\hbar^{2}}{8m\gamma_{p}^{2}}\nabla(\partial^{\beta}\mu^{\gamma}_{p}\cdot\partial^{\beta}\mu^{\gamma}_{p})
-\nabla(q_{p}\varphi+\frac{1}{2}m\textbf{v}_{p}^{2}) ,\end{equation}
where $\varphi$ is the scalar potential of electromagnetic field.

Equations keep they structure at account of the Darwin and annihilation interactions, but explicit form of $\widehat{\textbf{B}}$ and $\mbox{\boldmath $\chi$}$ changes. Consequently we find conservation of the generalized
quantum helicity $h_{a}$ associated with the $\mbox{\boldmath $\Omega$}_{(a),-}$, defined as $h_{(a),-}=\textbf{P}_{(a),-}\cdot\mbox{\boldmath $\Omega$}_{(a),-}$,
\begin{equation}\label{E-P_QP} \frac{d h_{(a),-}}{d t}=0\end{equation}

The rate of change of helicity $h_{(a),+}$ associated with the $\mbox{\boldmath $\Omega$}_{(a),+}$ is given by
\begin{equation}\label{E-P_QP} \frac{d h_{(a),+}}{d t}=\frac{2\hbar}{m\gamma_{a}}\mu^{\gamma}(\partial^{\beta} \hat{B}^{\gamma})\Omega_{(a),+}^{\beta}.\end{equation}

We also have that the rate of change of either
$h_{c}$ or $h_{q}$ is proportional to $d h_{(a),+}/(d t)$.

%\begin{equation}\label{E-P_QP} \end{equation}

\section{Conclusion}

We have developed the quantum hydrodynamics for electron-positron plasmas. This model requires to include the annihilation interaction. It gives contribution in the Euler equations and the magnetic moment evolution equations.

As consequence, contributions of the annihilation interaction in spectrum of plasma waves were found. We have found shifts of the eigen-frequencies of the transverse electromagnetic plane polarized waves, transverse spin-plasma waves, we have also included contribution of the quantum Bohm potential in dispersion of this branch of waves, longitudinal Langmuir waves. We do it for two limit cases of waves propagating parallel and perpendicular to the external magnetic field. We have also considered oblique propagation of longitudinal waves. It also contains contribution of the Fermi pressure, the Darwin interaction, and the quantum Bohm potential from the Euler equations.

We have obtained corresponding set of two non-linear Schrodinger equations. These equations are another representation of the set of QHD equations for eddy-free motion.

We have derived equations for the grand generalized
vorticities for each species. We have applied these equations to demonstrate existence of conserving helicity in electron-positron quantum plasmas of spinning particles with the Darwin and annihilation interactions.

We have generalized our model. We have derived the quantum kinetic equations from the many-particle Schrodinger equation. This kinetic theory is presented in the self-consistent field approximation. For simplicity of presentation we have shown interaction terms in the quasi-classic approximation. We applied the quantum kinetic equations to the Langmuir waves in absence of the external fields. We have found contribution of the Darwin interaction and spinless part of the annihilation interaction in frequency of the wave and in the collisionless Landau damping. These interactions  decrease the real and imaginary parts of spectrum.

Annihilation interaction may find its application in physics of quantum electron-ion plasmas. In this case the annihilation interaction reveals as result of virtual recombination of an electron and an ion, with subsequent ionization (since recombination is virtual).

These models open possibilities to consider different linear and non-linear quantum effects with full interaction between spinning electrons and positrons.

%%%%%%%%%%%%%%%%%%%%%%%%%%%%%%%%%%%%%%%%%%%%%

\begin{acknowledgements}
The author thanks Professor L. S. Kuz'menkov for fruitful discussions.
\end{acknowledgements}

\end{document}